\documentclass[12pt]{article}
\usepackage{amsmath}
\usepackage{amssymb}
\usepackage{amsfonts}
\usepackage{latexsym}
\usepackage{color}
\usepackage{graphicx}

\catcode `\@=11 \@addtoreset{equation}{section}
\def\theequation{\arabic{section}.\arabic{equation}}
\catcode `\@=12

\newcommand{\be}{\begin{equation}}
\newcommand{\en}{\end{equation}}
\newcommand{\bea}{\begin{eqnarray}}
\newcommand{\ena}{\end{eqnarray}}
\newcommand{\beano}{\begin{eqnarray*}}
\newcommand{\enano}{\end{eqnarray*}}
\newcommand{\bee}{\begin{enumerate}}
\newcommand{\ene}{\end{enumerate}}
\newcommand{\R}{\mathbb{R}}

\newcommand{\ST}{\mathcal{S}}

\newcommand{\Hil}{\mathcal{H}}

\newcommand{\Id}{1\!\!1}
\newcommand{\F}{\mathcal{F}}
\newcommand{\K}{\mathcal{K}}

\newcommand{\Nc}{\mathcal{N}}

\newcommand{\A}{\mathcal{A}}

\begin{document}
\title{An Operator--like Description of Love Affairs}

\author{{\Large Fabio Bagarello}\\
{\small Dipartimento di Metodi e Modelli Matematici}\\
{\small Facolt\`a di Ingegneria, Universit\`a di Palermo}\\
{\small Viale delle Scienze, I--90128  Palermo, Italy}\\
{\small e-mail: bagarell@unipa.it}\\
\vspace{2mm}\\
{\Large Francesco Oliveri}\\
{\small Dipartimento di Matematica, Universit\`a di Messina}\\
{\small Viale F. Stagno D'Alcontres 31, I--98166 Messina, Italy}\\
{\small email: oliveri@mat520.unime.it}}

\date{}
\maketitle
\begin{abstract}
\noindent We adopt the so--called \emph{occupation number representation}, originally used in
quantum mechanics and recently
considered in the description of stock markets, in the analysis of
the dynamics of love relations. We start with a simple model,
involving two actors (Alice and Bob): in the linear case we obtain  periodic dynamics,
whereas in the nonlinear regime either periodic or quasiperiodic solutions are found.
Then we extend the model to a love triangle involving Alice, Bob and a third actress,
Carla. Interesting features appear, and in particular we find analytical
conditions for the linear model of love triangle to have periodic or quasiperiodic
solutions. Numerical solutions are exhibited in the nonlinear case.
\end{abstract}

\vspace{2cm}

%{\bf PACS Numbers}:  .......

%\vfill

\newpage

% Section 1
\section{Introduction and preliminaries}
\label{sec:introduction} In a series of recent papers one of us
(F.B.) used the framework of quantum mechanics, operator algebra
and, in particular, of the so--called \emph{occupation number
representation} to discuss some toy models of stock markets
\cite{bag1,bag2,bag3,bag4}. The main motivation for such an
approach was that during the time evolution the main variables of
a closed stock market can take only discrete values. This feature
is well described by using the eigenvalues of some operators which
describe these variables, the so-called \emph{observables} of the
market. Moreover, the closed stock markets we have considered
admit conserved quantities, like the total number of shares or the
total amount of cash, and these conserved quantities are well
described in our framework. Here we want to show how the same
general approach could be used in dealing with a completely
different problem, \emph{i.e.}, the analysis of a \emph{love
affair}. In fact, it is  natural to measure the mutual affection
of the actors of our model using natural numbers (the higher the
number, the stronger the love), and to think that some conserved
quantities do exist in the game. It might be worth
recalling that sophisticated mathematical tools have been used
several times in the analysis of this problem, producing many
interesting results which can be found in
\cite{Strogatz,Strogatz:book,Rinaldi1998,Rinaldi,Sprott,Sprott1}, as well as
in an extensive monograph, \cite{wed}.
For instance, in \cite{Sprott}, some simple
dynamical models involving coupled ordinary differential
equations and describing the time variation of the love or hate
in a romantic relationship are considered. In particular, a linear model for
two individuals is discussed, and the extension to a love triangle, with nonlinearities causing chaotic dynamics, is also taken into account.

It should also be
considered that in the last few years a growing interest in
\emph{classical} application of \emph{quantum} ideas has appeared in the
literature. This involves very different fields like economics,
\cite{baa,scha}, biology, \cite{arndt}, and sociology,
\cite{aerts,aerts2} and references therein, just to cite a few, and is a strong encouragement to
carry on our analysis.

The paper is organized as follows. In Section
\ref{sec:firstmodel}, we consider a first simple model involving
two lovers, Alice and Bob, and we analyze the dynamics of their
relationship starting from  very natural technical assumptions. Both a
linear and a nonlinear model are considered; then, the equations
of motion are solved analytically (for the linear model) and
numerically (for the nonlinear model), under suitable (and fairly good)
approximations.

In Section \ref{sec:lovetriangle}, we consider a model in which
Bob has two relationships at the same time, and again we carry on
our dynamical analysis. Also in this case we  find an explicit
solution for the linear model, while the nonlinear one is
discussed numerically. Section \ref{sec:conclusions} contains our
conclusions, while, to keep the paper self--contained,
the Appendix reviews few useful facts in quantum mechanics and occupation
number representation.

\section{A first model}
\label{sec:firstmodel}

The first model we have in mind consists of a
couple of lovers, Bob and Alice, which mutually \emph{interact} exhibiting
a certain \emph{interest} for each other. Of course, there are
several degrees of possible interest, and to a given Bob's
interest  for Alice (LoA, \emph{level of attraction}) there
corresponds a related reaction (\emph{i.e.}, a different LoA) of
Alice for Bob. Now, let us see how this mechanism could be
described in terms of creation and annihilation operators.

Extending what has been done in \cite{bag1,bag2,bag3,bag4}, we now
introduce $a_1$ and $a_2$,  two independent bosonic operators.
This means that they obey the commutation rules \be
[a_i,a^\dagger_k]=a_i\,a^\dagger_k-a^\dagger_k\,a_i=\Id\,\delta_{i,k},
\label{21} \en while all the other commutators are trivial:
$[a_i,a_k]=[a^\dagger_i,a^\dagger_k]=0$, for all $i$ and $k$.
Further, let $\varphi_0^{(j)}$ be the \emph{vacuum} of $a_j$,
$a_j\,\varphi_0^{(j)}=0$, $j=1,2$. By using $\varphi_0^{(j)}$ and
the operators $a_j^\dagger$, we may construct the following
vectors: \be
\varphi_{n_j}^{(j)}:=\frac{1}{\sqrt{n_j!}}\,(a_j^\dagger)^{n_j}\,\varphi_0^{(j)},\qquad
\varphi_{n_1,n_2}:=\varphi_{n_1}^{(1)}\otimes\varphi_{n_2}^{(2)},
\label{22} \en where $n_j=0,1,2,\ldots$, and $j=1,2$. Let us also
define $N_j=a_j^\dagger\,a_j$, $j=1,2$, and $N=N_1+N_2$. Hence
(see the Appendix), for all $j$,
\beano
 \left\{
    \begin{array}{ll}
N_j\varphi_{n_1,n_2}=n_j\varphi_{n_1,n_2}, \\
N_j\,a_j\,\varphi_{n_1,n_2}=(n_j-1)\,a_j\varphi_{n_1,n_2},\\
N_j\,a_j^\dagger\,\varphi_{n_1,n_2}=(n_j+1)\,a_j^\dagger\varphi_{n_1,n_2}.
\end{array}
        \right. \enano

Then, we also have
$N\varphi_{n_1,n_2}=(n_1+n_2)\varphi_{n_1,n_2}$. As usual, the
Hilbert space $\Hil$ in which the operators live is obtained by
taking the closure of the linear span of all these vectors, for
$n_j\geq0$, $j=1,2$. A state over the system  is  a normalized
linear functional $\omega_{n_1,n_2}$ labeled by two \emph{quantum
numbers} $n_1$ and $n_2$ such that
$\omega_{n_1,n_2}(x)=\left<\varphi_{n_1,n_2},x\,\varphi_{n_1,n_2}\right>$,
where $\left<.,.\right>$ is the scalar product in $\Hil$ and $x$
is an arbitrary operator on $\Hil$.

In this paper we associate the (integer) eigenvalue  $n_1$ of
$N_1$ to the LoA that Bob experiences for Alice: the higher the
value of $n_1$ the more Bob desires Alice. For instance, if
$n_1=0$, Bob just does not care about Alice. We use $n_2$, the
eigenvalue of $N_2$, to label the attraction of Alice for Bob. A
well known (surely simplified) law of attraction stated in our
language says that if $n_1$ increases then $n_2$ decreases and
viceversa\footnote{This law has inspired many (not only) Italian love songs
over the years!}. Following the same general strategy as in
\cite{bag1,bag2,bag3,bag4}, this suggests to use the following
self--adjoint operator to describe the dynamics (see Appendix) of
the relationship: \be
H=\lambda\left(a_1^{M_1}\,{a_2^\dagger}^{M_2}+\hbox{h.c.}\right).
\en Here, $\hbox{h.c.}$ stands for hermitian conjugate, and $M_1$
and $M_2$ give a measure of the kind of mutual reaction between
Bob and Alice: if $M_1$ is large compared to $M_2$, Bob will
change his status very fast compared with Alice. The opposite
change is expected for $M_2$ much larger than $M_1$, while, for
$M_1$ close to $M_2$, they will react essentially
with the same speed. These claims will be justified in the rest of the paper.
However, as it is already clear at this stage, it is enough to introduce a single index $M$ rather
than $M_1$ and $M_2$, since $M$ plays the role of a {\em relative behavior}.
For this reason we will choose the hamiltonian \be
H=\lambda\left(a_1^{M}{a_2^\dagger}+\hbox{h.c.}\right),
\label{23}\en where $\lambda$ is the interaction parameter (which
could also be seen as a time scaling parameter). The physical
meaning of $H$ can be deduced considering the action of, say,
$a_1^{M}\,{a_2^\dagger}$ on the  vector describing the system at
time $t=0$, $\varphi_{n_1,n_2}$. This means that, at $t=0$, Bob is
in the state $n_1$, i.e. $n_1$ is Bob's LoA, while Alice is in the
state $n_2$. However, because of the definition of
$\varphi_{n_1,n_2}$, $a_1^{M}\,{a_2^\dagger}\,\varphi_{n_1,n_2}$,
which is different from zero only if $M<n_1$,
is proportional to $\varphi_{n_1-M,n_2+1}$. Hence, Bob's interest for Alice decreases of $M$  units
while Alice's interest for Bob increases of 1 unit. Of course, the Hamiltonian (\ref{23}) also
contains the opposite behavior. Indeed, because of the presence of $a_2\,{a_1^\dagger}^M$ in $H$, if
$n_2\geq1$ we see that
$a_2\,{a_1^\dagger}^M\varphi_{n_1,n_2}$ is proportional to
$\varphi_{n_1+M,n_2-1}$: hence, Bob's interest is increasing (of
$M$ units) while Alice looses interest in Bob. It is not hard to
check that $I(t):=N_1(t)+M\, N_2(t)$ is a constant of motion:
$I(t)=I(0)=N_1(0)+M\, N_2(0)$, for all $t\in\R$. This is a
consequence of the following commutation result: $[H,I]=0$.
Therefore, during the time evolution, a certain {\em global
attraction} is preserved and can only be exchanged between Alice
and Bob: notice that this reproduces our original point of view
on the love relation between Alice and Bob.

Now we have all the ingredients to derive the equations of motion
for our model. These are found by assuming, as we have already done
implicitly,  the same Heisenberg--like dynamics which
works perfectly for quantum systems, and which is natural in the present
operatorial settings. Of course, this is a strong
assumption and should be checked \emph{a posteriori}, finding the
dynamical behavior deduced in this way and showing that this gives
reasonable results. More explicitly, we are assuming that the time
evolution $X(t)$ of a given observable $X$ of the system is given
by $X(t)=e^{iHt}Xe^{-iHt}$. The equations of motion for the number
operators $N_1(t)$ and $N_2(t)$, which are needed to deduce the
\emph{rules of the attraction}, turn out to be the following: \bea
 \left\{
    \begin{array}{ll}
 \dot N_1(t)=i\,\lambda M\left(a_2^\dagger(t)(a_1(t))^M-a_2(t)(a_1(t)^\dagger)^M\right),\\
\dot N_2(t)=-i\,\lambda
\left(a_2^\dagger(t)(a_1(t))^M-a_2(t)(a_1(t)^\dagger)^M\right).
\end{array}
        \right.\label{24} \ena

By using this system, it is straightforward to check directly that
$I(t)$ does not depend on time. However, this system is not
closed, so that it may be more convenient to replace (\ref{24}) by
the differential system for the annihilation operators $a_1(t)$
and $a_2(t)$: \bea
 \left\{
    \begin{array}{ll} \dot a_1(t)=-i\,\lambda M\,a_2(t)(a_1(t)^\dagger)^{M-1}\\
\dot a_2(t)=-i\,\lambda\,(a_1(t))^M.\end{array}\right. \label{25}
\ena Then, we may use the solutions of this system to construct
$N_j(t)=a_j^\dagger(t)\,a_j(t)$, $j=1,2$. Equations (\ref{25}),
together with their adjoint, produce a closed system. Of course,
there exists a simple situation for which the system (\ref{25})
can be exactly solved quite easily: $M=1$. In this case, which
corresponds to the assumption that Alice and Bob react with the
same speed, system (\ref{25}) is already closed and the solution
is easily found: \be a_1(t)=a_1\cos(\lambda t)-ia_2\sin(\lambda
t), \qquad a_2(t)=a_2\cos(\lambda t)-i a_1 \sin(\lambda t). \en
Now, if we assume that at $t=0$ Bob and Alice are respectively in
the $n_1$'th and $n_2$'th LoA's, the state of the system at $t=0$
is $\omega_{n_1,n_2}$ (see Appendix).  Therefore, calling
$n_j(t):=\omega_{n_1,n_2}(N_j(t))$, $j=1,2$, we find that
\be\label{n1lin2} n_1(t)=n_1\cos^2(\lambda t)+n_2\sin^2(\lambda
t),\qquad n_2(t)=n_2\cos^2(\lambda t)+n_1\sin^2(\lambda t), \en so
that, in particular, $\omega_{n_1,n_2}(I(t))=n_1+n_2$, as
expected. The conclusion is quite simple and close to our view of
how the law of the attraction works: the infatuations of Alice and
Bob oscillate in such a way that when Bob's LoA increases, that of
Alice decreases and viceversa, with a period which is directly
related to the value of the interaction parameter $\lambda$. In
particular, as  is natural, setting $\lambda=0$ in equation
(\ref{n1lin2}), implies that both Alice and Bob stay in their
initial LoA's. The solution in (\ref{n1lin2}) justifies, in a sense, our approach: the law of the attraction between Alice and
Bob is all contained in a single operator, the hamiltonian of the
model, whose explicit expression can be easily deduced using
rather general arguments. The related dynamics is exactly the one
we expected to find. Hence, the use of the Heisenberg equations of
motion seems to be justified, at least for this simple model. A similar result was found for the linear model also in \cite{Sprott}, under suitable assumptions about the parameters.

Much harder is the situation when $M>1$. In this case we do need
to consider the adjoint of (\ref{25}) to close the system and,
nevertheless, an exact solution can not be obtained. However, it
is possible to generate a numerical scheme to find solutions of
our problem, and this is the content of the next subsection.

\subsection{Numerical results for $M>1$}
\label{sec:firstmodel_sub1} The first remark is that, as stated
above, the Hilbert space of our theory, $\Hil$, is
infinite--dimensional. This means that both Bob and Alice may experience
infinite different LoA's. This makes the situation rather hard
from a computational point of view and, furthermore, looks like a
useless difficulty. As a matter of fact, it is enough to assume
that Bob (respectively, Alice) may pass through $L_1$
(respectively, $L_2$) different LoA's ($L_1$ and $L_2$ fixed
positive integers), which efficiently describe their mutual
attraction. This means that our effective Hilbert space,
$\Hil_{eff}$, is finite--dimensional and is generated by the
orthonormal basis \be \F=\{\varphi_{\underline
n}:=\varphi_{n_1,n_2}, \quad
n_j=0,1,\ldots,L_j,\,j=1,2\}=\{\varphi_{\underline n}, \quad
\underline n \in\K\}, \en with obvious notation. Hence,
$dim(\Hil_{eff})=(L_1+1)(L_2+1)$ is exactly the cardinality of
$\K$.

Calling $\Id_{eff}$ the identity operator over
$\Hil_{eff}$ the closure relation for $\F$ looks like
$\sum_{\underline n\in K}|\varphi_{\underline
n}><\varphi_{\underline n}|=\Id_{eff}$. It is a standard exercise
in quantum mechanics to check that \bea
 \left\{
    \begin{array}{ll}
 (a_1)_{\underline k,\underline k'}:=<\varphi_{\underline k},a_1\varphi_{\underline k'}>=\sqrt{k_1'}\,\delta_{k_1,k_1'-1}\,\delta_{k_2,k_2'},\\
 (a_2)_{\underline k,\underline k'}:=<\varphi_{\underline k},a_2\varphi_{\underline k'}>=\sqrt{k_2'}\,\delta_{k_2,k_2'-1}\,\delta_{k_1,k_1'},
 \end{array}
        \right.\label{2728} \ena
so that, by using the resolution of the identity in $\Hil_{eff}$,
we can produce a \emph{projected} expression of the operators
$a_j$ as follows: \be
\begin{aligned}
&a_1\rightarrow A_1:=\sum_{\underline k,\underline k'\in
K}\,<\varphi_{\underline k},a_1\varphi_{\underline
k'}>|\varphi_{\underline k}><\varphi_{\underline k'}|=
\sum_{\underline k'\in K}\,\sqrt{k_1'}\,|\varphi_{k_1'-1,k_2'}><\varphi_{k_1'k_2'}|,\\
&a_2\rightarrow A_2:=\sum_{\underline k,\underline k'\in
K}\,<\varphi_{\underline k},a_2\varphi_{\underline
k'}>|\varphi_{\underline k}><\varphi_{\underline k'}|=
\sum_{\underline k'\in
K}\,\sqrt{k_2'}\,|\varphi_{k_1',k_2'-1}><\varphi_{k_1'k_2'}|.
\end{aligned}
\label{29210} \en In other words, while the $a_j$'s act on $\Hil$,
the related matrices $A_j$'s act on the finite dimensional Hilbert
space $\Hil_{eff}$. This means that, while the $a_j(t)$'s are
unbounded operators (which could be represented as infinite
matrices), the $A_j(t)$'s are  $(L_1+1)(L_2+1)\times
(L_1+1)(L_2+1)$ matrices whose elements, for $t=0$, can be deduced
by (\ref{29210}). Thus, these are really bounded operators and, as
we will see, they all have the following property: for all choices
of $L_j$ a certain power $R_j$ exists such that $A_j^{R_j}=0$,
$(A_j^\dagger)^{R_j}=0$, condition which is not shared by the
original $a_j$. This fact is related to the approximation which
produces $\Hil_{eff}$ out of the original $\Hil$. Now, system (\ref{25}) can be written in a
formally identical way as \bea
 \left\{
    \begin{array}{ll}
\dot A_1(t)=-i\,\lambda M\,A_2(t)(A_1(t)^\dagger)^{M-1},\\
\dot A_2(t)=-i\,\lambda\,(A_1(t))^M,
 \end{array}
        \right.\label{211} \ena where $A_1(t)$ and $A_2(t)$ are bounded matrices
rather than unbounded operators. It should be stressed that the
change $a_j\rightarrow A_j$ does not destroy the existence of an
integral of motion, which is clearly $
I_{eff}(t)=\Nc_1(t)+M\Nc_2(t)$, where $\Nc_j(t)=A_j(t)^\dagger
A_j(t)$.

The numerical scheme is quite simple: we just have to fix the
dimensionality of $\Hil_{eff}$, that is $L_1$ and $L_2$
(\emph{i.e.}, the number of LoA's), the value of $M$ in
(\ref{23}), and the matrices $A_j(0)$. Then, by choosing a
reliable scheme for integrating numerically a set of ordinary
differential equations, we may construct a solution for the
equations (\ref{211}) in a prescribed time interval.
Fron now on, we take $L_1=L_2=K$ ($K$ suitable positive integer), whereupon
$dim(\Hil_{eff})=(K+1)^2$, and we consider the orthonormal basis of
$\Hil_{eff}$ in the following order (the order is important to fix
the form of the matrix!):
$$
\F=\left\{\varphi_{00},\varphi_{01},\ldots,\varphi_{0K},\varphi_{10},\varphi_{11},\ldots,\varphi_{1K},\ldots,\varphi_{K0},\varphi_{K1},\ldots,\varphi_{KK}\right\}.
$$
Then we get
$$
\begin{aligned}
A_1(0)&=\left(
\begin{array}{cccccc}
\mathbf{0}_K & \mathbf{0}_K & \ldots & \ldots & \mathbf{0}_K &\mathbf{0}_K   \\
\Id_K & \mathbf{0}_K & \ldots & \ldots & \mathbf{0}_K & \mathbf{0}_K   \\
\mathbf{0}_K & \sqrt{2}\Id_K & \mathbf{0}_K &\ldots & \mathbf{0}_K &\mathbf{0}_K   \\
\mathbf{0}_K & \mathbf{0}_K & \sqrt{3}\Id_K & \ldots & \mathbf{0}_K &\mathbf{0}_K\\
\ldots & \ldots & \ldots & \ldots & \ldots & \ldots \\
\ldots & \ldots & \ldots & \ldots & \ldots & \ldots \\
\mathbf{0}_K & \mathbf{0}_K & \ldots & \ldots &\sqrt{K-1}\Id_K & \mathbf{0}_K
\end{array}
\right), \\
A_2(0)&=\left(
\begin{array}{ccccc}
\underline x & \mathbf{0}_K & \ldots & \ldots & \mathbf{0}_K   \\
\mathbf{0}_K & \underline x & \ldots & \ldots &\mathbf{0}_K   \\
\ldots & \ldots & \ldots &\ldots & \ldots \\
\ldots & \ldots & \ldots &\ldots & \ldots \\
\mathbf{0}_K & \mathbf{0}_K & \ldots & \ldots &\underline{x}   \\
\end{array}
\right),
\end{aligned}
$$
where $\mathbf{0}_K$ and $\Id_K$ are the null and the identity matrices of order $K$
respectively, whereas
\[
\underline{x}=\left(
\begin{array}{ccccc}
 0 &  0 &  0 &\ldots & 0   \\
 1 &  0 &  0 & \ldots & 0  \\
 0 &   \sqrt{2} & 0  \ldots & 0 \\
 \ldots & \ldots & \ldots & \ldots & \ldots \\
 0 & 0 &\ldots &\sqrt{K-1} & 0
\end{array}
\right).
\]

It is easy to verify that $A_1(0)^{K-1}\neq \mathbf{0}_K$ and $A_2(0)^{K-1}\neq \mathbf{0}_K$,
while $A_1(0)^K=A_2(0)^K=\mathbf{0}_K$. This simply means that,
since because of our original approximation
$\Hil\rightarrow\Hil_{eff}$, there are only $K+1$ different levels;
then, if we try to act more than $K+1$ times on a certain state,
the only effect we get is  just to annihilate that state. In other
words, we can not move Bob or Alice to a $-1$ or $K+1$ LoA  by acting
with $A_j$ or $A_j^\dagger$ since these states do not exist!
However, this argument does not apply to the operators $a_j$ in $\Hil$,
since in the original Hilbert space there is no upper bound to the value
of the LoA's of the two lovers.

Suppose now that for $t=0$   the system is in the state
$\omega_{\underline n}$, where $\underline n=(n_1,n_2)$. This
means, as usual, that $\underline n$ describes the LoA of Alice
and Bob. As we have done explicitly for $M=1$,  if we want to know
how Bob's LoA varies with time, we have to compute
\[
n_1(t):=<\varphi_{\underline n},\Nc_1(t)\varphi_{\underline n}>=
\|A_1(t)\varphi_{\underline n}\|^2=\sum_{\underline k\in
K}|(A_1(t))_{\underline k,\underline n}(t)|^2,
\]
in terms of the matrix elements. Analogously, to compute how
Alice's LoA varies with time, we need to compute
\[
n_2(t):=<\varphi_{\underline n},\Nc_2(t)\varphi_{\underline n}>=
\|A_2(t)\varphi_{\underline n}\|^2=\sum_{\underline k\in
K}|(A_2(t))_{\underline k,\underline n}(t)|^2.
\]

The system (\ref{211}) is numerically integrated  for different choices of $K$
by taking $M=2$ and $\lambda=0.1$. We use a variable order Adams--Bashforth--Moulton PECE solver \cite{ode113} as implemented in MATLAB$^\circledR$'s ode113 routine.

\begin{figure}
\begin{center}
\includegraphics[width=0.7\textwidth]{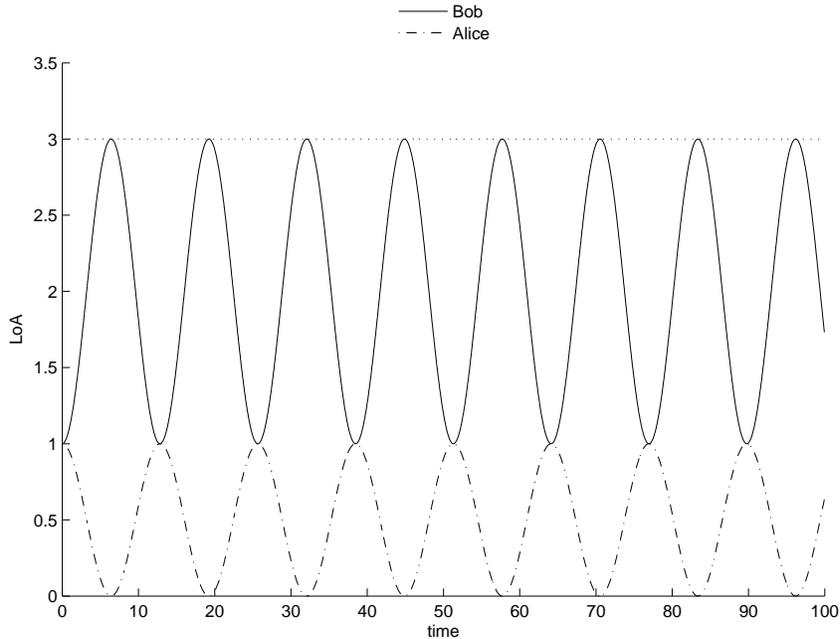}
\caption{\label{fig1}\footnotesize $K=3$, $M=2$: Alice's and Bob's
LoA's vs. time with initial condition $(1,1)$. A periodic behavior is observed.}
\end{center}
\end{figure}

In Figure~\ref{fig1},  the time evolutions of Alice's and
Bob's LoA's in the case $K=3$ with equal initial conditions $n_1(0)=n_2(0)=1$ are displayed.
Two clear oscillations in opposite phase can be observed: Alice and Bob react
simultaneously but \emph{in different directions}.
This is in agreement with our naive point of view of Alice--Bob's love
relationship, and confirms that the Heisenberg equations of motion
can really be used to describe the dynamics of this classical
system, also in this nonlinear (and likely more realistic) case.
The horizontal dotted line on the top in Figure~\ref{fig1} (and also in Figures~\ref{fig2} and \ref{fig3} below) represents the integral of motion $I_{eff}(t)$;
in some sense, it provides a check of the quality of the numerical solution.

\begin{figure}
\begin{center}
\includegraphics[width=0.47\textwidth]{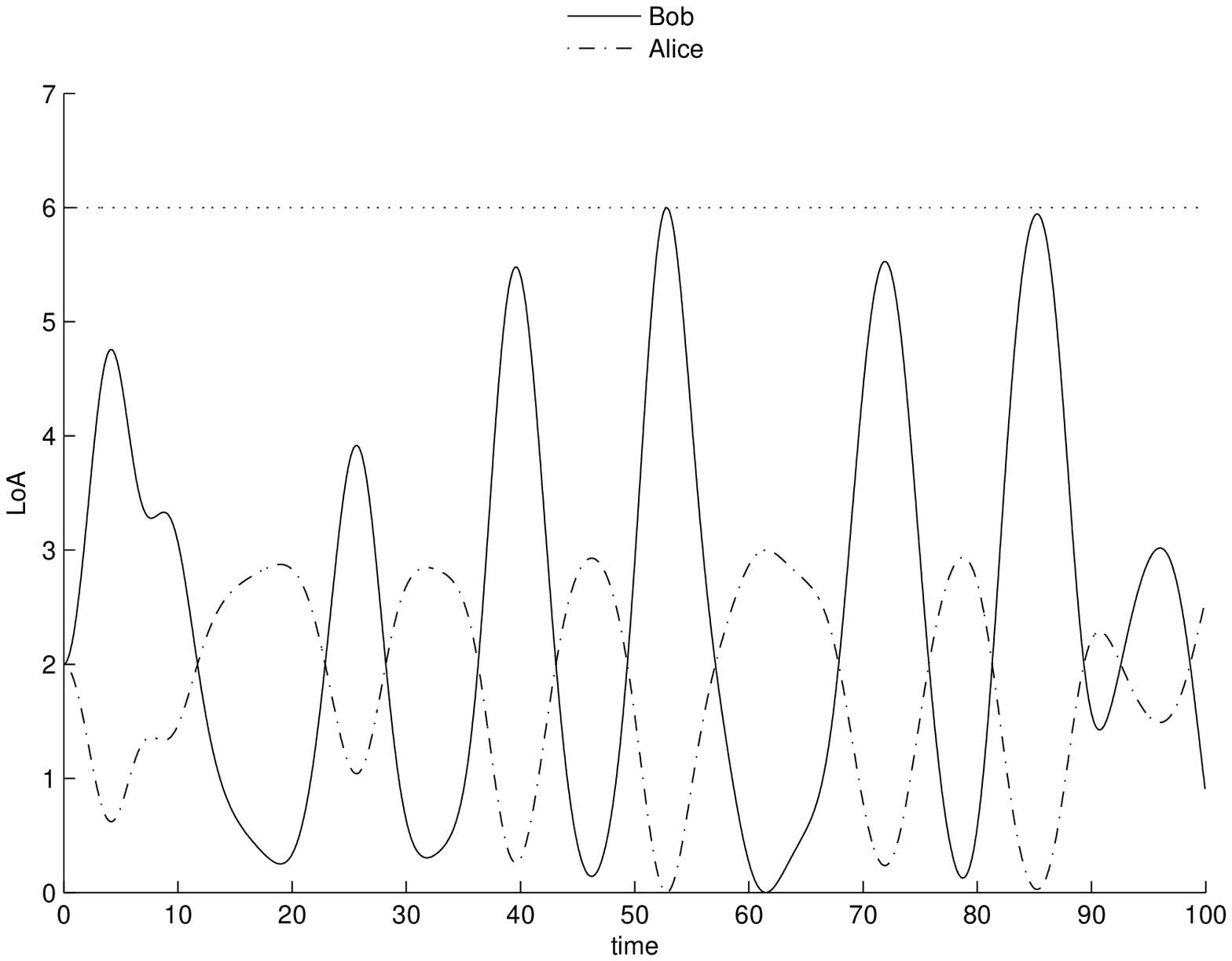}\hspace{8mm}
\includegraphics[width=0.47\textwidth] {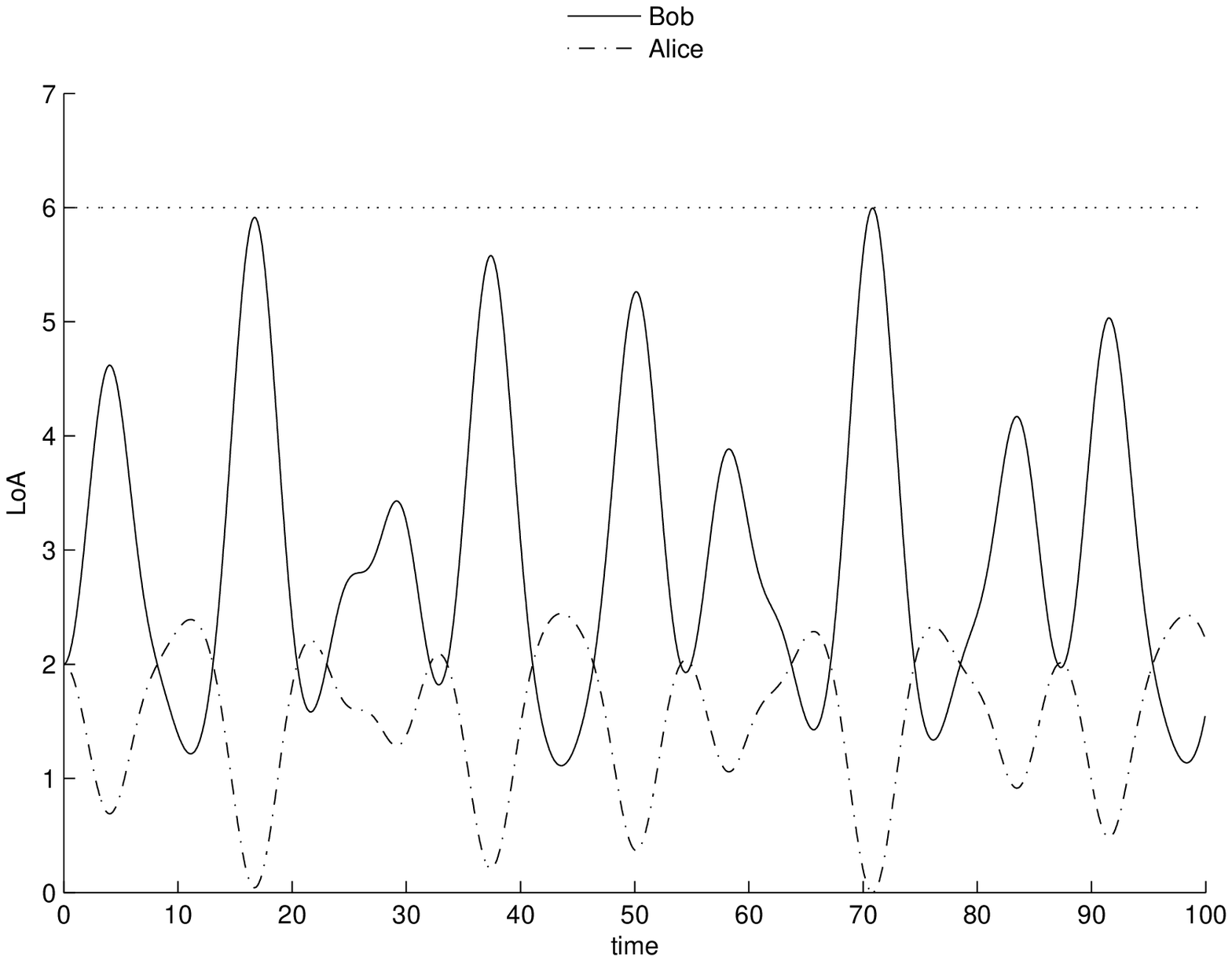}\hfill\\
\includegraphics[width=0.47\textwidth]{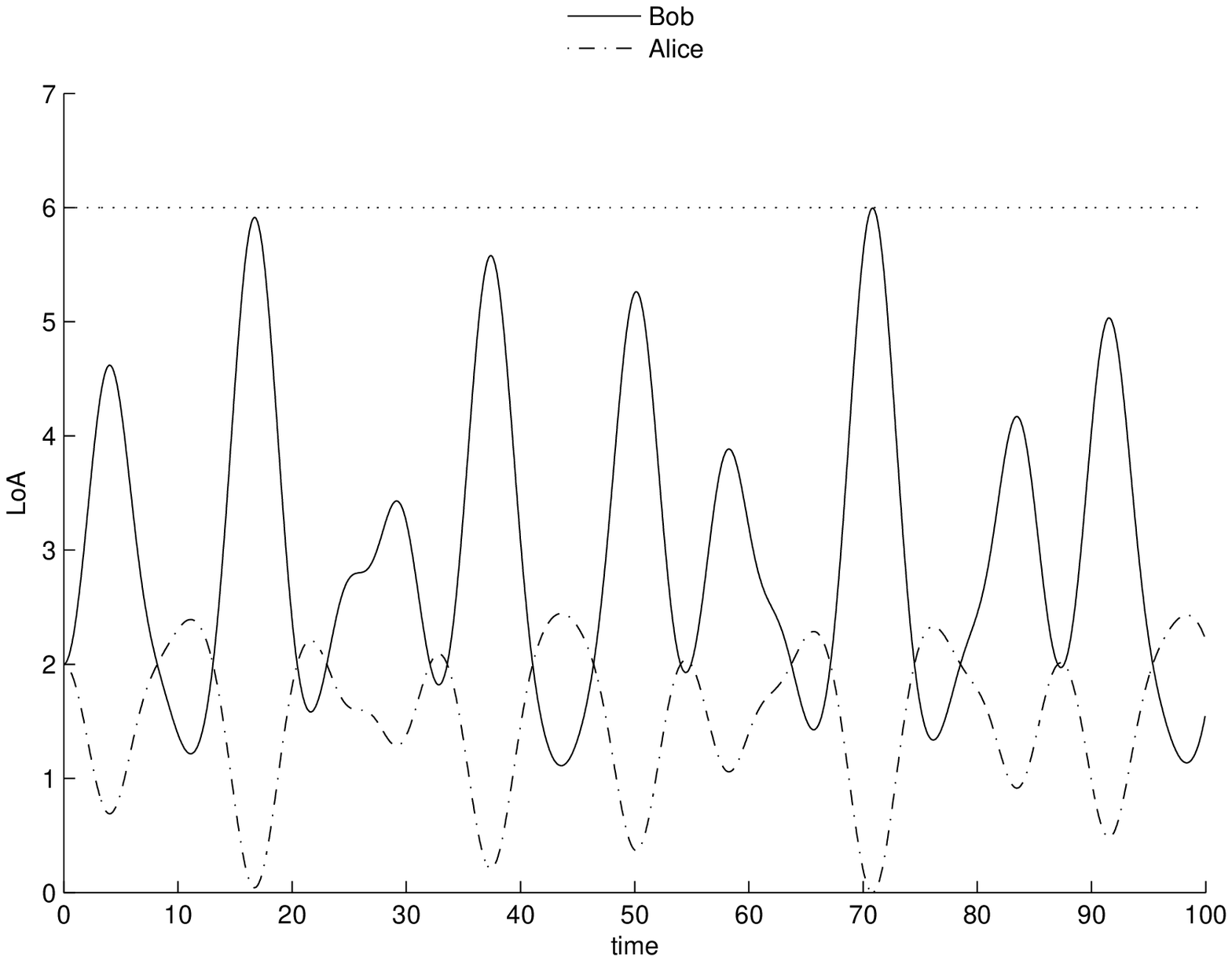}\hspace{8mm}
\includegraphics[width=0.47\textwidth] {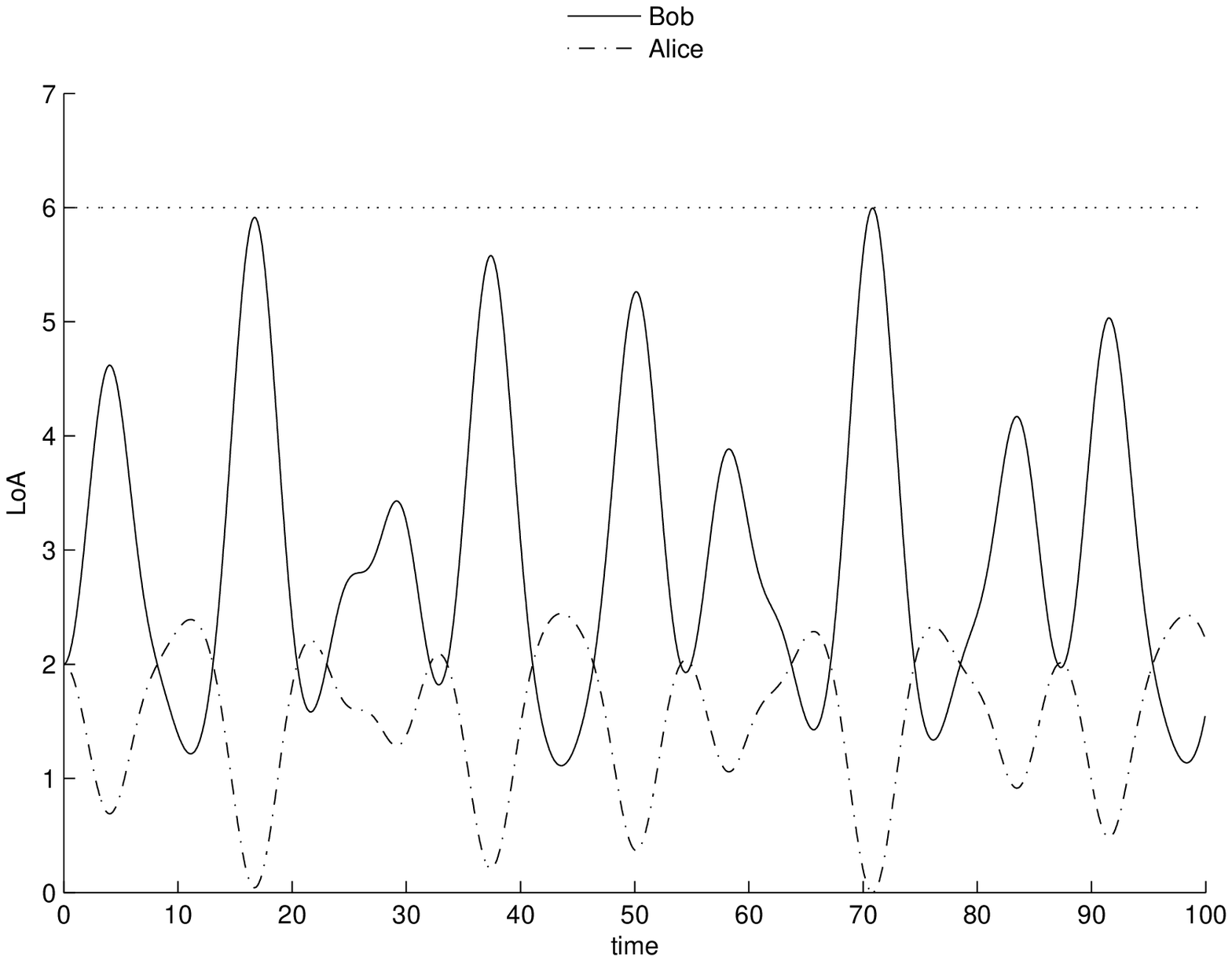}\hfill\\
\caption{\label{fig2}\footnotesize Alice's and Bob's
LoA's vs. time with initial conditions $(2,2)$ and $M=2$: $K=3$ (top left), $K=6$ (top right),
$K=7$ (bottom left), $K=8$ (bottom right): $K=6$ is already a good approximation.}
\end{center}
\end{figure}

In Figure~\ref{fig2}, the evolution of Alice's and Bob's LoA's with
initial conditions $(2,2)$ for different dimensions of $\Hil_{eff}$ is plotted. As it can
be observed, when $K=3$, the values of LoA's go beyond the maximum admissible value ($K$); this
suggests to improve the approximation, {\em i.e.} to increase the dimension of $\Hil_{eff}$. By
taking $K=6$, we find
that Alice's and Bob's LoA's assume values within the bounds. Moreover, by performing
the numerical integration with higher values of $K$ ($K=7$ and $K=8$),
the same dynamics as when $K=6$  is recovered. We also see that, as in Figure~\ref{fig1}, $n_1(t)$
increases when $n_2(t)$ decreases, and viceversa.

The same main features are found in Figure~\ref{fig3}, where the initial condition $(1,3)$ and
different values of $K$ ($K=5, 6, 7, 8$) are considered. The choice $K=5$ turns out to be a poor
approximation; in fact, the values of LoA's can not be described in its $\Hil_{eff}$. On the
contrary, already for $K=6$, the related $\Hil_{eff}$ turns out to be a fairly good substitute of
$\Hil$,
since the values of LoA's do not exceed the limits during the time evolution. Furthermore, the
dynamics is \emph{stable} for increasing values of $K$.

\begin{figure}
\begin{center}
\includegraphics[width=0.47\textwidth]{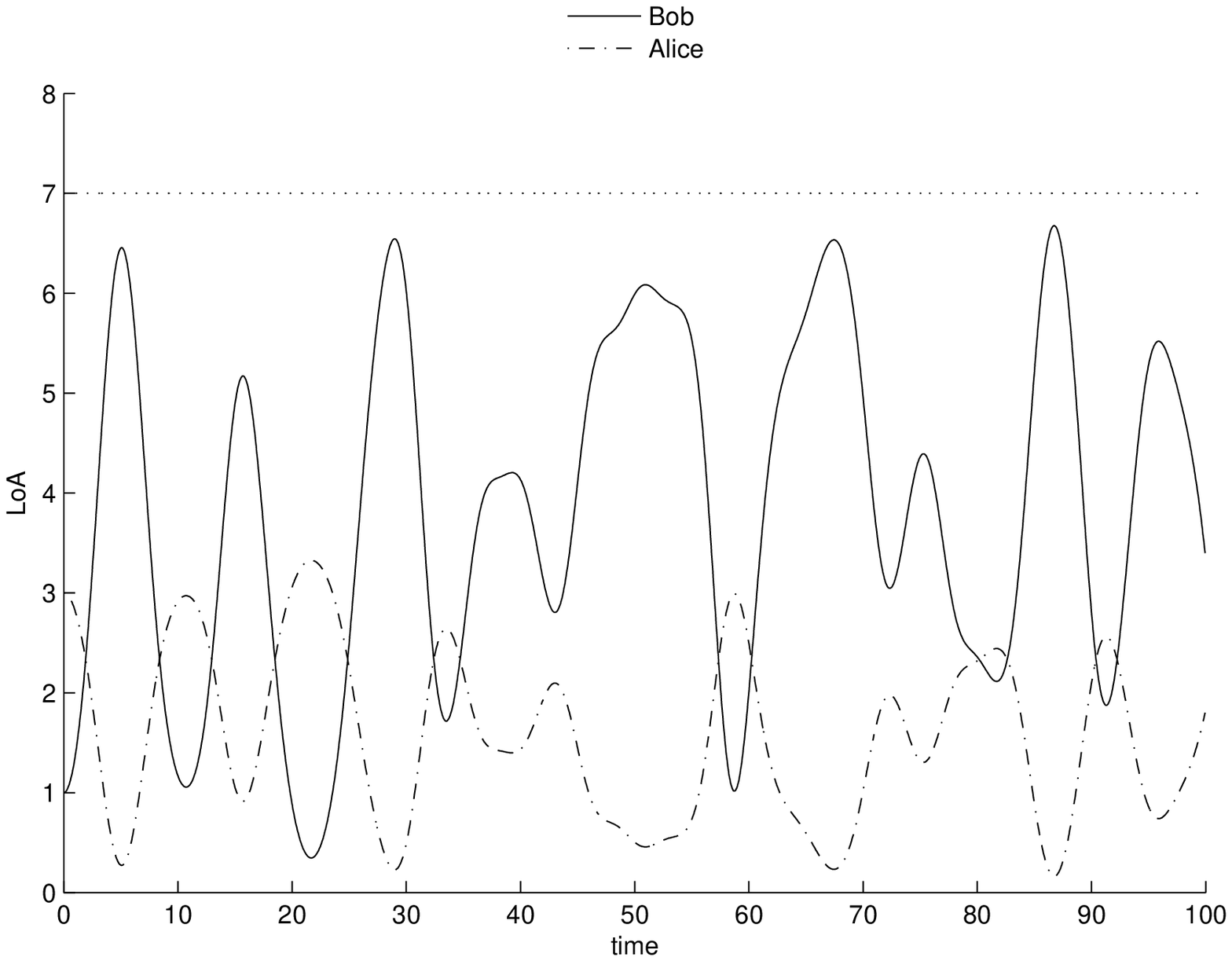}\hspace{8mm}
\includegraphics[width=0.47\textwidth]{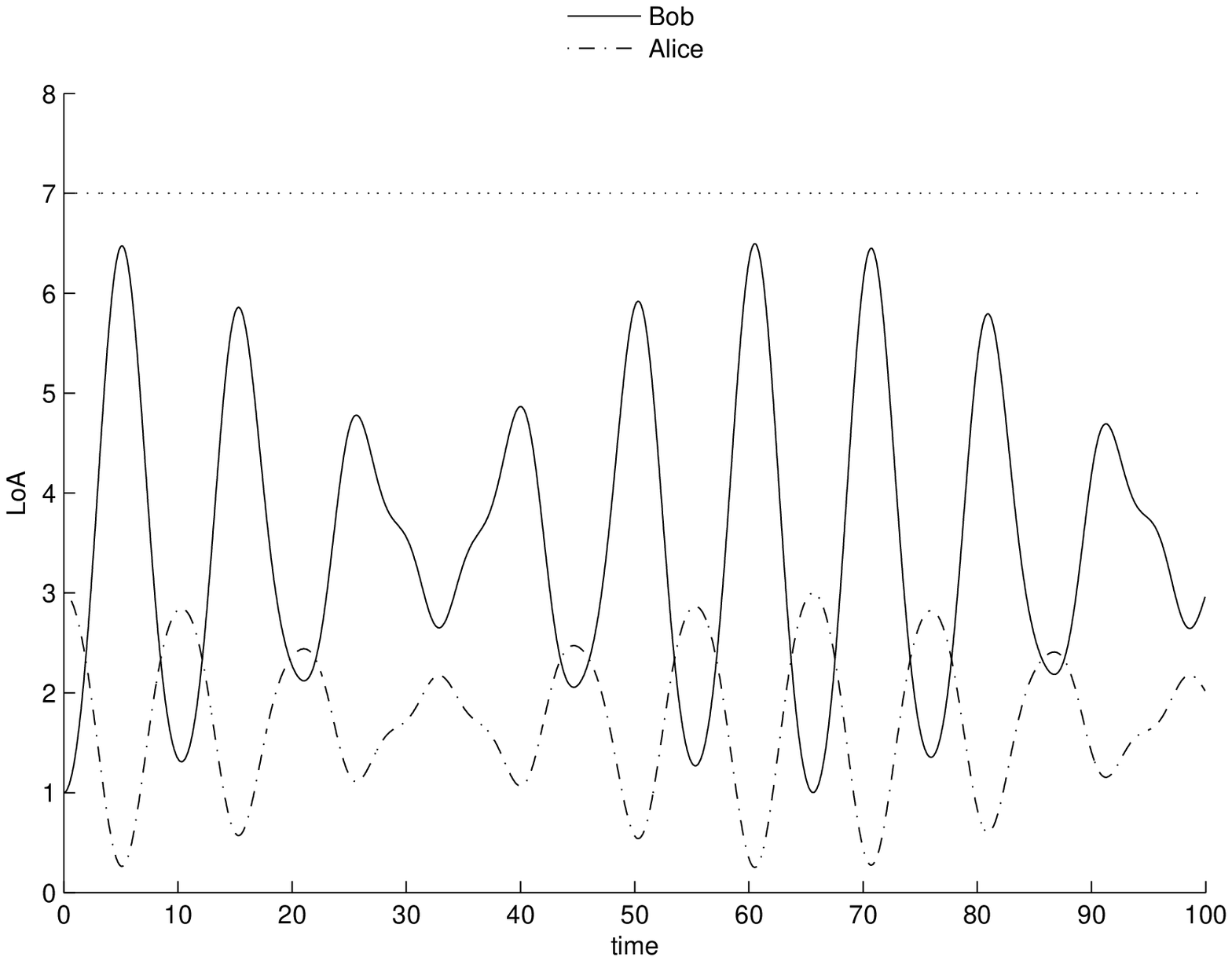}\\
\includegraphics[width=0.47\textwidth]{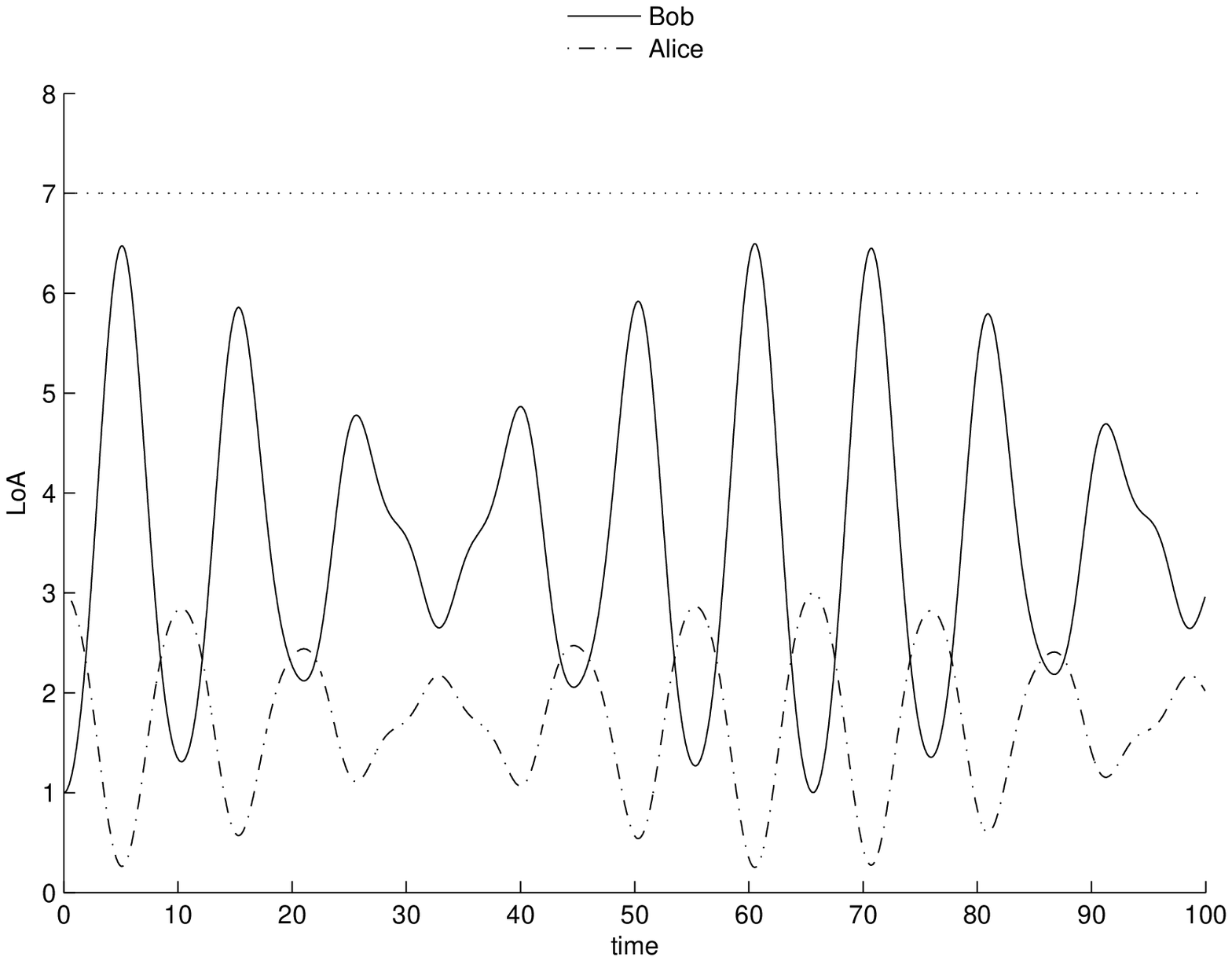}\hspace{8mm}
\includegraphics[width=0.47\textwidth]{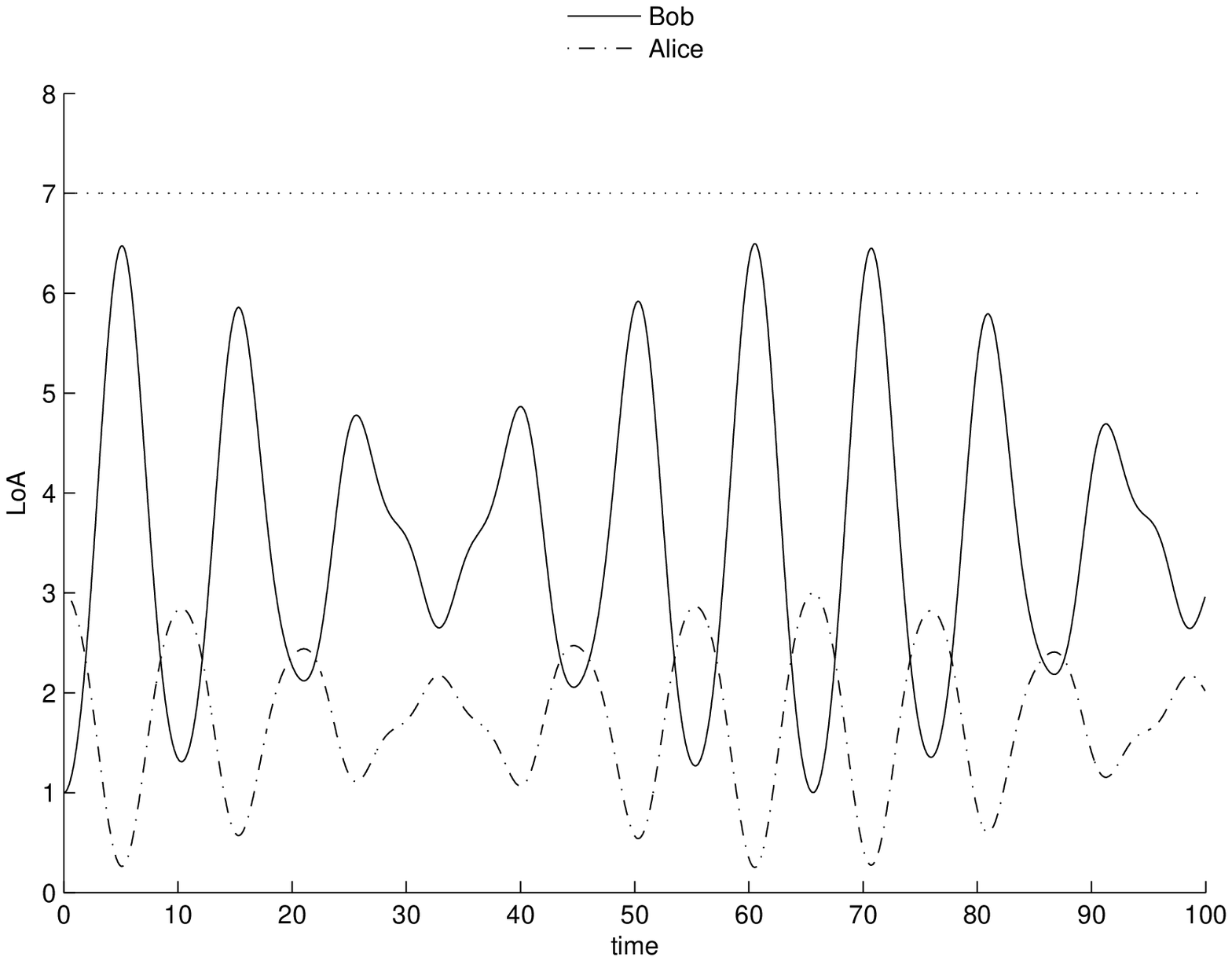}\\
\caption{\label{fig3}\footnotesize Alice's and Bob's
LoA's vs. time with initial condition $(1,3)$ and $M=2$: $K=5$ (top left), $K=6$ (top right),
$K=7$ (bottom left), $K=8$ (bottom right): again, $K=6$ is already a good approximation.}
\end{center}
\end{figure}

Let us summarize the situation: as a consequence of our approximation,
$\Hil\rightarrow\Hil_{eff}$, we need to replace $I(t)$ with
$I_{eff}(t)=\Nc_1(t)+M\,\Nc_2(t)$. As already noticed, because of
(\ref{211}), $\dot I_{eff}(t)=0$, \emph{i.e.}, $I_{eff}(t)$ is a constant of
motion for the approximate model. Hence, it  happens that if
$\Nc_2(t)$ decreases during its time evolution, $\Nc_1(t)$ must
increase since $I_{eff}(t)$ has to stay constant. In this way, for
some value of $t$, it may happen that $\Nc_1(t)>K$ (see Figure~\ref{fig2} for $K=3$, and
Figure~\ref{fig3} for $K=5$). This problem is cured simply by fixing higher values of
$K$, as numerically shown in Figures~\ref{fig2} and \ref{fig3}, where the
choice $K=6$ allows us to capture the right dynamics, that remains unchanged also
when $K=7$ and $K=8$. In a sense, the dimension of $\Hil_{eff}$
can be \emph{a priori} fixed looking at the value of the integral of motion. In fact, if
for some $\overline{t}$, $\Nc_2(\overline{t})=0$, then $\Nc_1(\overline{t})$
is equal to $I_{eff}(0)$, which in turn depends on the initial conditions. Therefore,
the dimension of $\Hil_{eff}$ must be greater than or equal to $I_{eff}(0)$; last, but not least,
our numerical tests allow us to conjecture that, increasing
the dimension of $\Hil_{eff}$, the resulting dynamics is not affected.

It may be worth stressing that this feature only appears because of the
numerical approach we are adopting here, and nothing has to
do with our general framework. This is clear, for instance,
considering the solution of the linear situation where no
approximation is needed and no effect like this is observed.
\begin{figure}
\begin{center}
\includegraphics[width=0.47\textwidth]{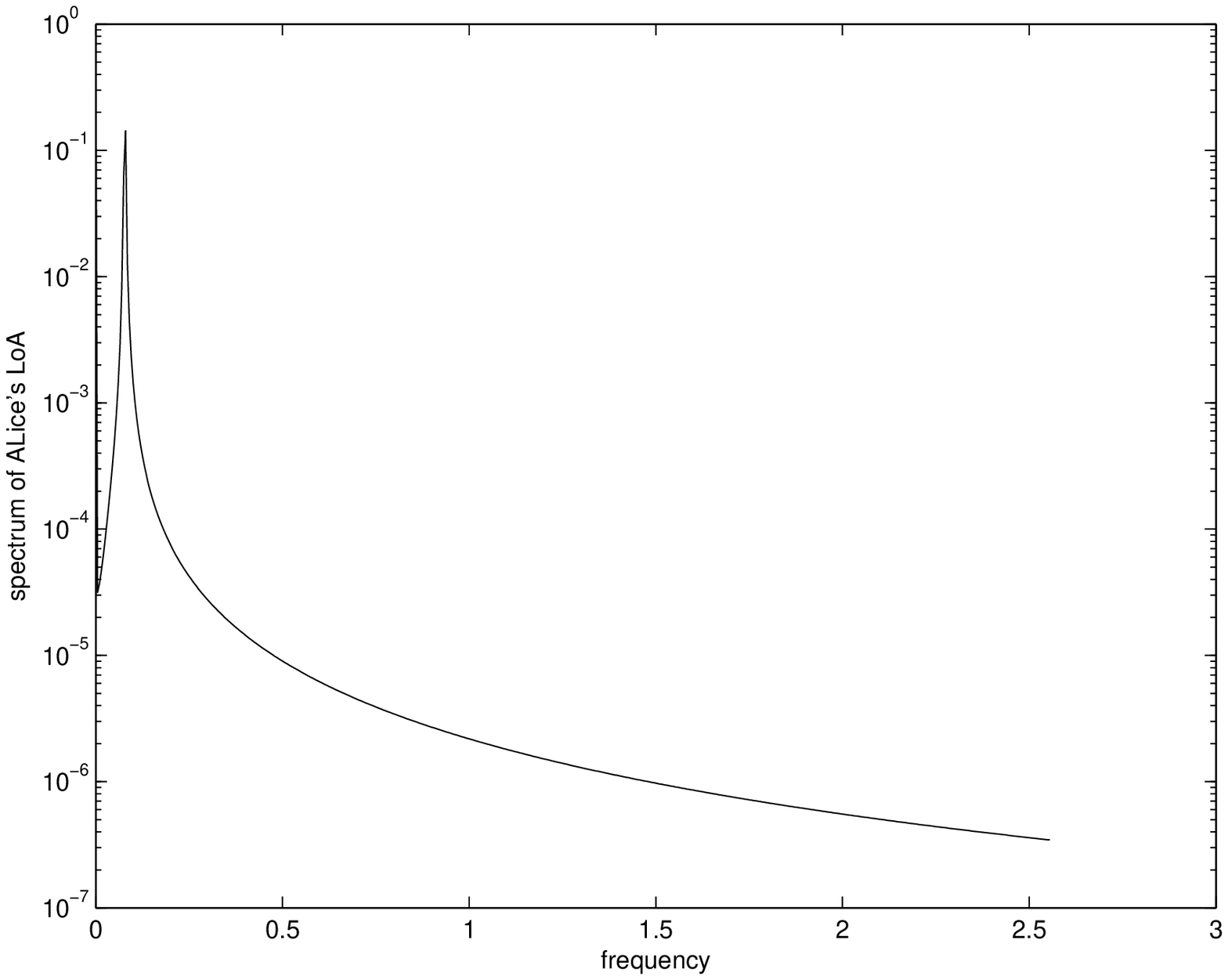}\hspace{8mm}
\includegraphics[width=0.47\textwidth]{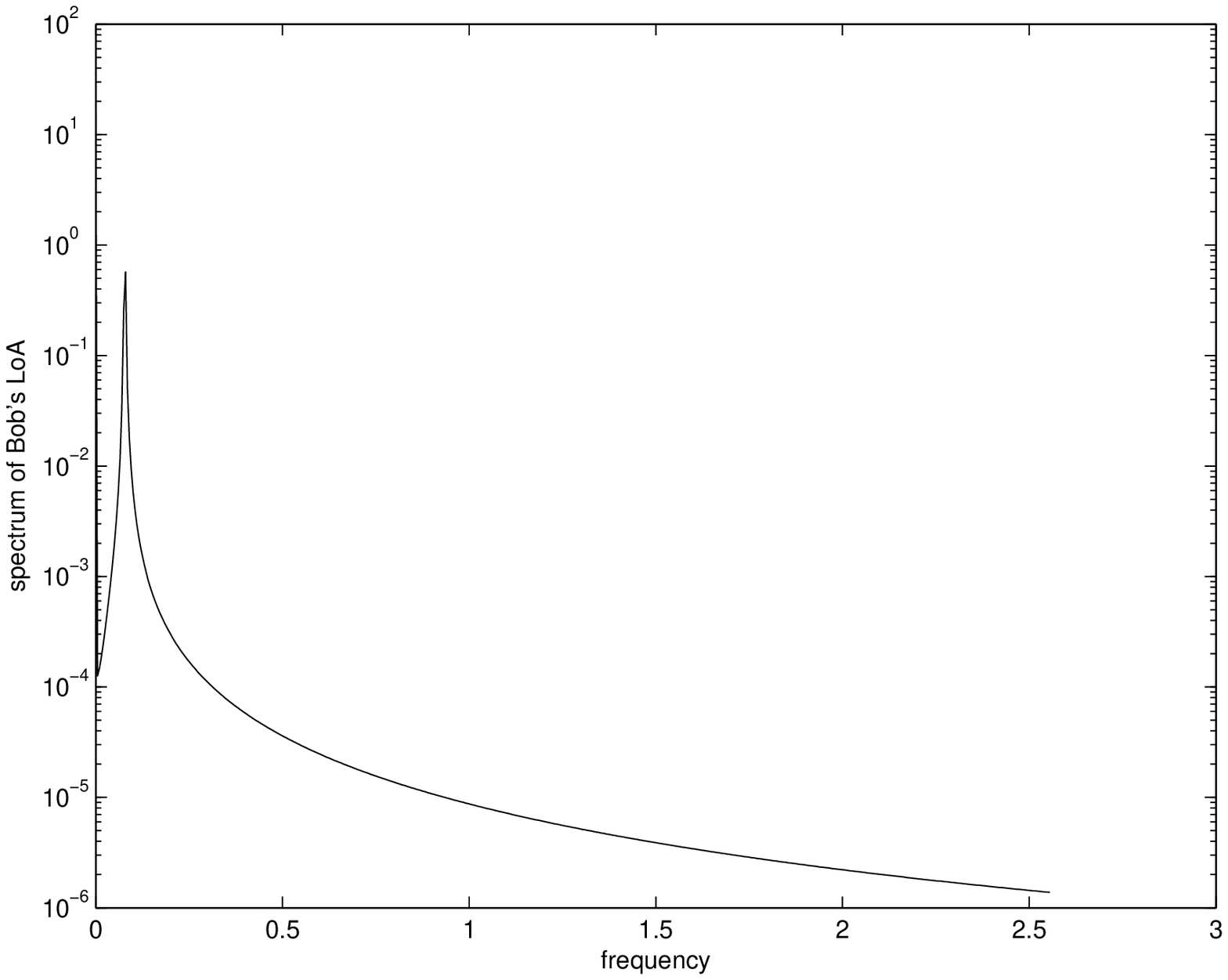}\\
\includegraphics[width=0.47\textwidth]{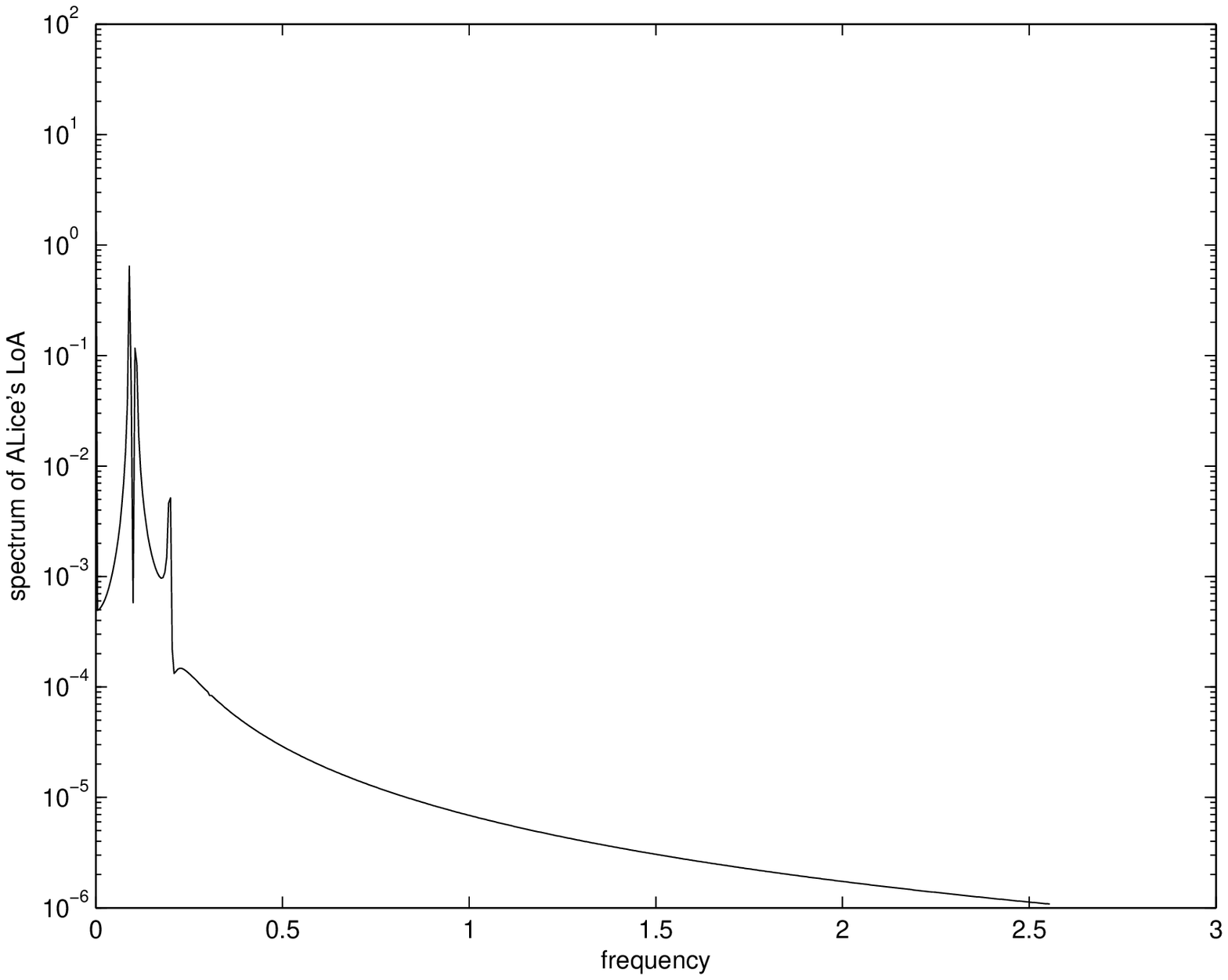}\hspace{8mm}
\includegraphics[width=0.47\textwidth]{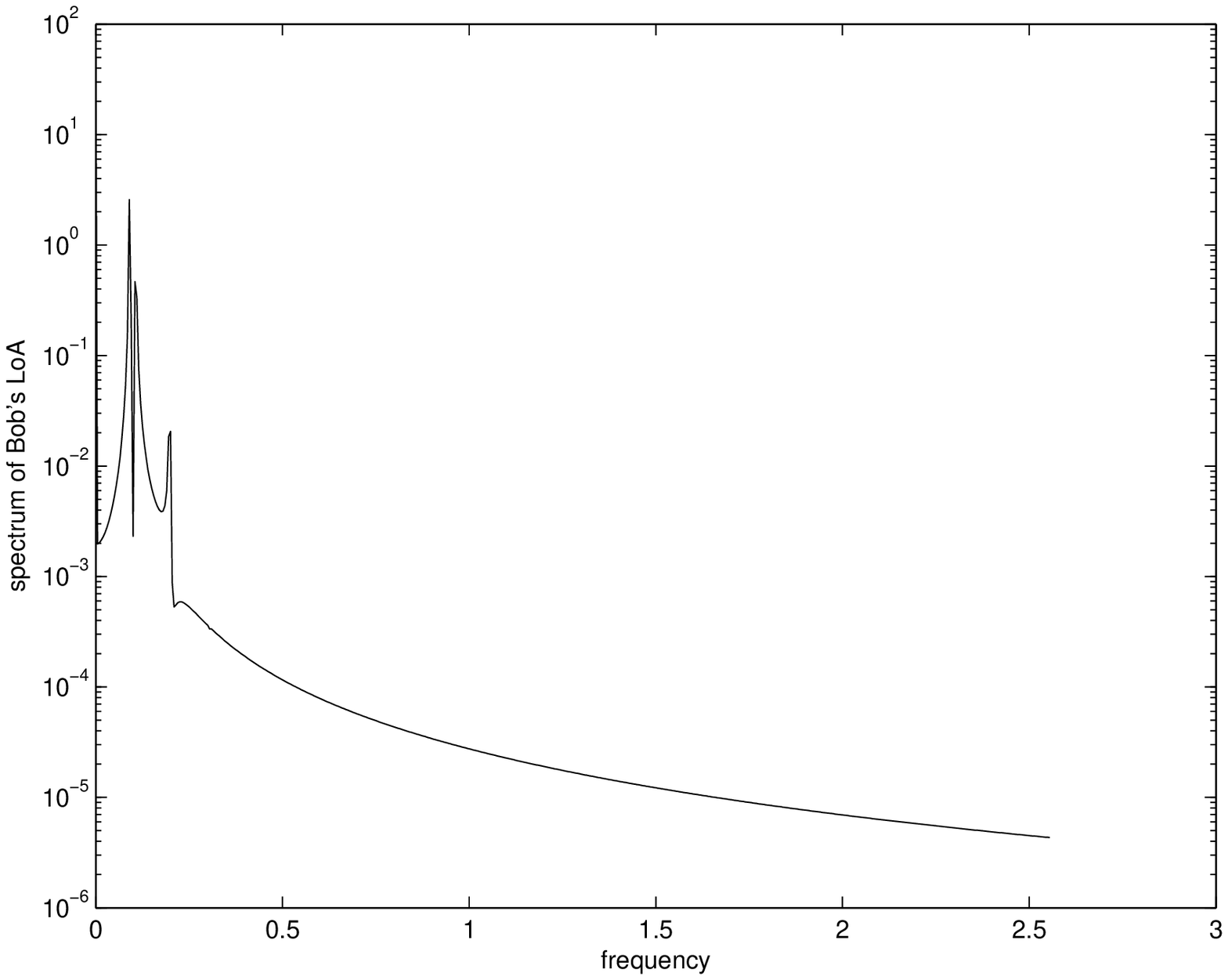}\\
\caption{\label{fig4}\footnotesize Power spectra vs. frequency of numerical solutions for
Alice's LoA (left) and Bob's LoA (right) for initial conditions
$(1,1)$ (top: periodic dynamics) and $(1,3)$ (bottom: quasiperiodic dynamics).}
\end{center}
\end{figure}

In this nonlinear case, the numerical results seem to show that a periodic motion is
not recovered for all initial conditions, contrarily to what analytically proved in the case
$M=1$. In  Figures~\ref{fig2} and \ref{fig3},
rather than a simple periodic time evolution (like that displayed in Figure~\ref{fig1}),
a quasiperiodic behavior seems to emerge, as the plots of the power spectra of
the time series representing the numerical solutions (see Figure~\ref{fig4}, bottom) also suggest.
However, what emerges from our numerical tests is that, once the dimension of $\Hil_{eff}$ is sufficient to capture the dynamics, only periodic or quasiperiodic solutions are obtained. 
This does not exclude that, for initial conditions requiring dimensions of the Hilbert space higher than those considered here, a richer dynamics (say, chaotic) may arise. 
Further numerical investigations in this direction are in progress.

\section{A generalization: a love triangle}
\label{sec:lovetriangle} In this section we will generalize our
previous model by inserting a third ingredient. Just to fix the
ideas, we will assume that this third ingredient, Carla, is Bob's
lover, and we will use the same technique to describe the
interactions among the three. The situation can be summarized as
follows:
\begin{enumerate}
\item Bob can interact with both Alice and Carla, but Alice
(respectively, Carla) does not suspect of Carla's (respectively,
Alice's) role in Bob's life;

\item if Bob's LoA for Alice increases then Alice's LoA for Bob
decreases and viceversa;

\item analogously, if Bob's LoA for Carla increases then Carla's
LoA for Bob decreases and viceversa;

\item if Bob's LoA for Alice increases then his LoA for Carla
decreases (not necessarily by the same amount) and viceversa.

\end{enumerate}

In order to simplify the computations, we assume  for the moment
that the \emph{action and the reaction} of the lowers have the
same strength. Therefore, repeating our previous considerations,
we take at first $M=1$. The Hamiltonian of the system is
a simple generalization of that in (\ref{23}) with $M=1$: \be
H=\lambda_{12}\left(a_{12}^\dagger\,a_2+a_{12}\,a_2^\dagger\right)+\lambda_{13}\left(a_{13}^\dagger\,a_3+a_{13}\,a_3^\dagger\right)+
\lambda_{1}\left(a_{12}^\dagger\,a_{13}+a_{12}\,a_{13}^\dagger\right).
\label{31} \en Here the indices 1, 2 and 3 stand for Bob, Alice
and Carla, respectively, the $\lambda_{\alpha}$'s are real
coefficients measuring the relative interaction strengths and the different
$a_{\alpha}$, $\alpha=12, 13, 2, 3$, are bosonic operators, such
that $[a_\alpha,a_\beta^\dagger]=\delta_{\alpha,\beta}\Id$, while
all the other commutators are zero. The three terms in $H$ are
respectively related to points 2., 3. and 4. of the above list. As
usual, we also introduce some number operators:
$N_{12}=a^\dagger_{12}\,a_{12}$, describing  Bob's LoA for Alice,
$N_{13}=a^\dagger_{13}\,a_{13}$, describing Bob's LoA  for Carla,
$N_{2}=a^\dagger_{2}\,a_{2}$, describing Alice's LoA  for Bob and
$N_{3}=a^\dagger_{3}\,a_{3}$, describing Carla's LoA for Bob. If
we define $J:=N_{12}+N_{13}+N_2+N_3$, which represents the total
level of LoA of the triangle, this is a conserved quantity:
$J(t)=J(0)$, since $[H,J]=0$. It is also possible to check that
$[H,N_{12}+N_{13}]\neq 0$, so that the total Bob's LoA is not
conserved during the time evolution.

The equations of motion for the variables $a_\alpha$'s can be
deduced as usual and we find: \bea\left\{
\begin{array}{ll} i\,\dot a_{12}(t)=\lambda_{12} \,a_2(t)+\lambda_1\,a_{13}(t),\\
i\,\dot a_{13}(t)=\lambda_{13} \,a_3(t)+\lambda_1\,a_{12}(t),\\
i\,\dot a_2(t)=\lambda_{12}\,a_{12}(t),\\
i\,\dot a_3(t)=\lambda_{13}\,a_{13}(t).
\end{array} \right.
\label{32} \ena This system can be explicitly solved and the
solution can be written as
\be
A(t)=U^{-1}\exp\left(-i\Lambda_d\,t\right)\,U\,A(0), \label{33}
\en
where $A$ is the (column) vector with components
$(a_{12},a_{13},a_{2},a_{3})$, $\Lambda_d$ is the diagonal matrix
of the eigenvalues of the matrix
\[
\Lambda=\left(
  \begin{array}{cccc}
    0 & \lambda_1 & \lambda_{12} & 0 \\
    \lambda_1 & 0 & 0 & \lambda_{13} \\
    \lambda_{12} & 0 & 0 & 0 \\
    0 & \lambda_{13} & 0 & 0 \\
  \end{array}
\right),
\]
and $U$ is the matrix which diagonalizes $\Lambda$. The
eigenvalues of the matrix $\Lambda$ are solutions of the
characteristic polynomial
\be
\label{charpoly}
\lambda^4-(\lambda_1^2+\lambda_{12}^2+\lambda_{13}^2)\lambda^2+\lambda_{12}^2\lambda_{13}^2=0,
\en
\emph{i.e.},
\be
\lambda_{\pm,\pm}=\pm\sqrt{(\lambda_1^2+\lambda_{12}^2+\lambda_{13}^2)\pm
\sqrt{(\lambda_1^2+\lambda_{12}^2+\lambda_{13}^2)^2-4\lambda_{12}^2\lambda_{13}^2}}.
\en
Since it is trivially
\be
\frac{\lambda_{+,\pm}}{\lambda_{-,\pm}}=-1,
\en
we have that the commensurability of all the eigenvalues is guaranteed if and only if
\be
\frac{\lambda_{+,+}}{\lambda_{-,-}}=\frac{p}{q},
\en
with $p$ and $q$ nonvanishing positive integers, that is, if and only if the condition
\be
\label{33bis}
\frac{\lambda_1^2+\lambda_{12}^2+\lambda_{13}^2}{\lambda_{12}\lambda_{13}}=\frac{p}{q}+\frac{q}{p}
\en
holds true.

Therefore, by computing $N_{12}(t)$,  $N_{13}(t)$,  $N_{2}(t)$ and
$N_{3}(t)$, the solutions are in general quasiperiodic with two periods, and become periodic if
the condition (\ref{33bis}) is satisfied.

\begin{figure}
\begin{center}
\includegraphics[width=0.47\textwidth]{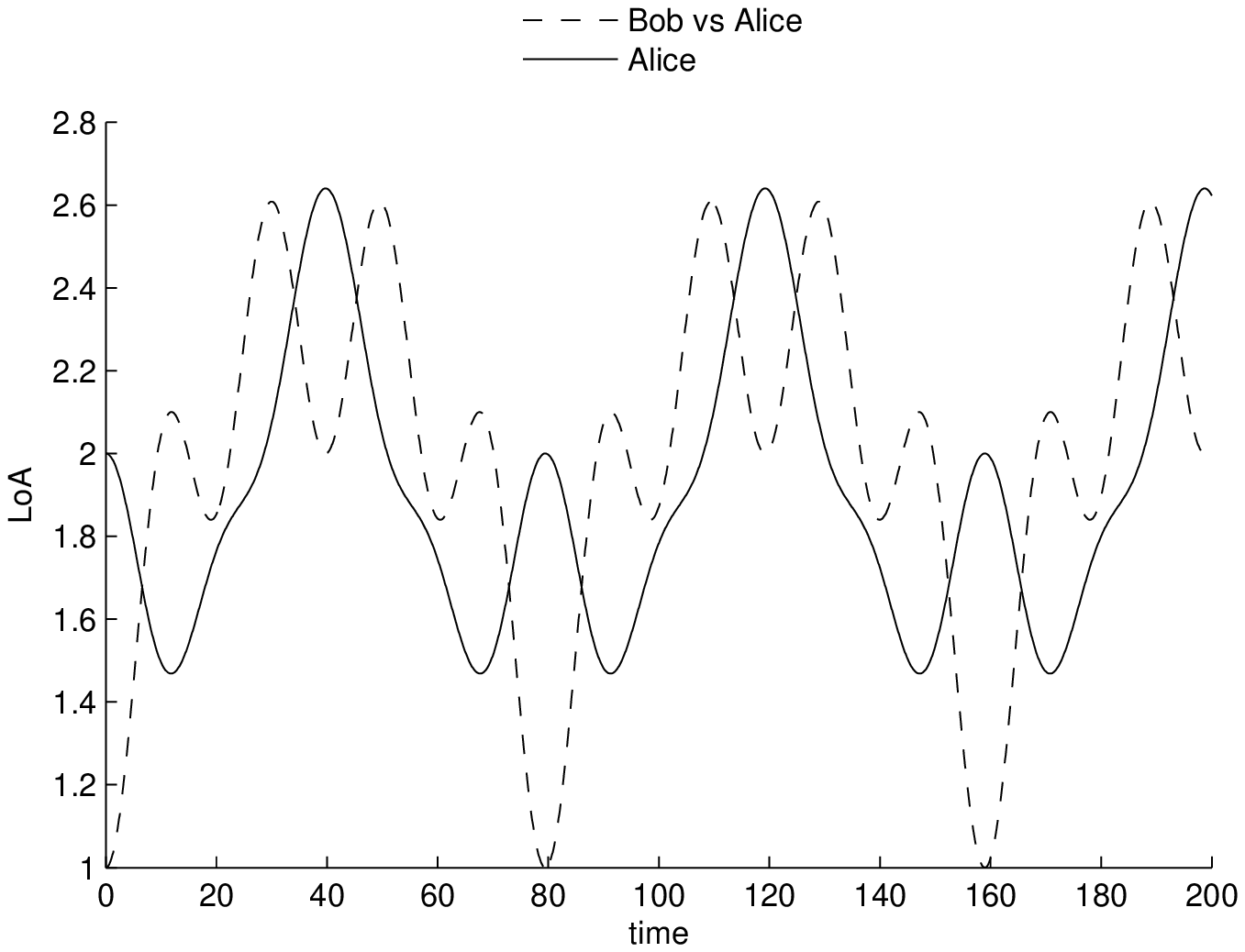}\hspace{8mm}
\includegraphics[width=0.47\textwidth] {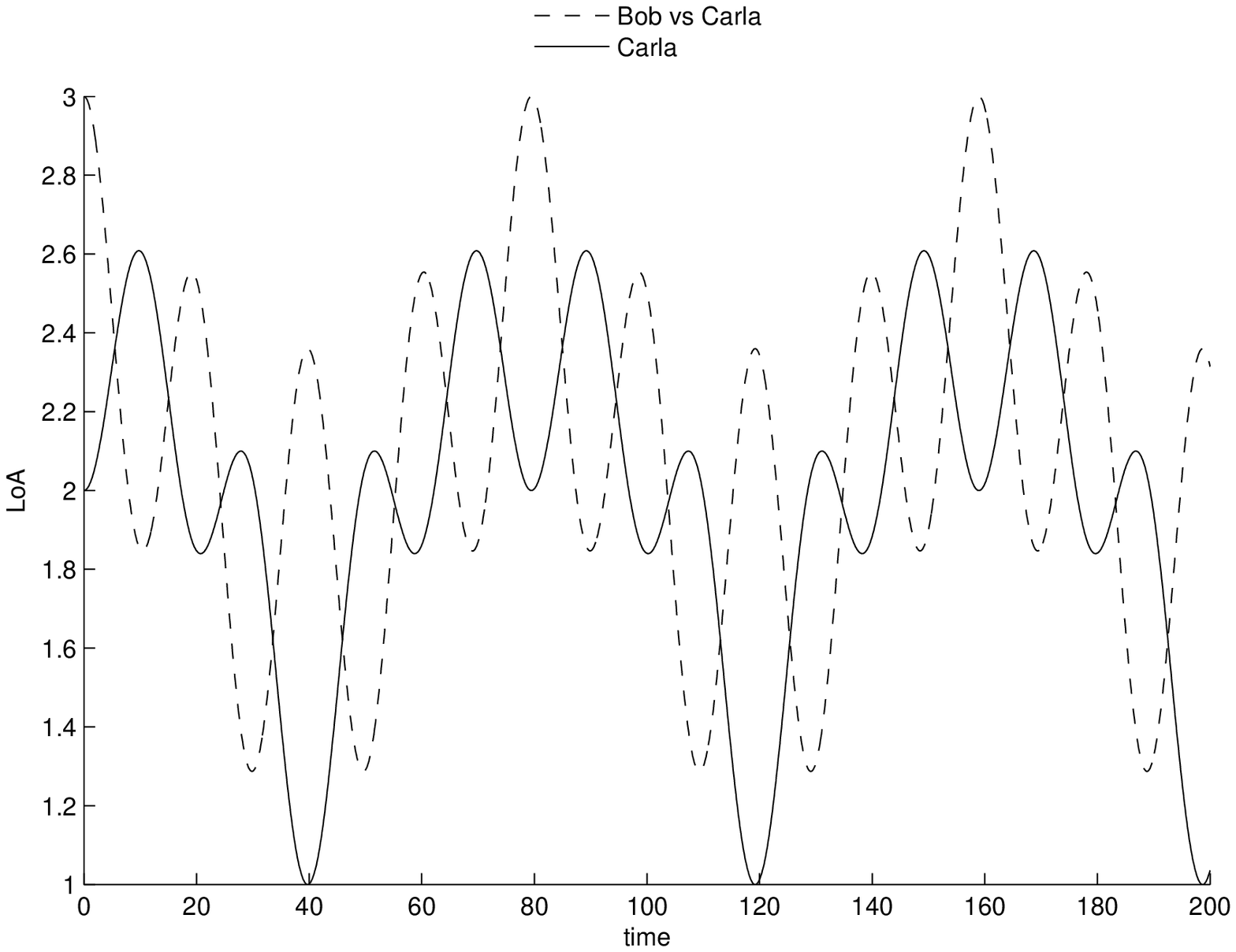}\hfill\\
\caption{\label{fig6}\footnotesize ($L_1=L_2=3$: Alice's and Bob's
LoA's vs. time with initial condition $(0,2)$ (left) and Carla's
and Bob's LoA's vs. time with initial condition $(2,3)$ (right). Periodic behaviours are observed.}
\end{center}
\end{figure}

In Figure~\ref{fig6}, we plot the solutions in the periodic case for
$L_1=L_2=3$, $n_{12}(0)=0$, $n_{13}(0)=3$, $n_2(0)=n_3(0)=2$: Bob
is strongly attracted by Carla, while both Carla and Alice
experience the same LoA with respect to Bob. The parameters
involved looks like these: $\lambda_{12}=\frac{1}{10}$,
$\lambda_{13}=\frac{1}{8}$ and $\lambda_{1}=\frac{3}{40}$, which
satisfies (\ref{33bis}) for $p=2$ and $q=1$.

\begin{figure}
\begin{center}
\includegraphics[width=0.47\textwidth]{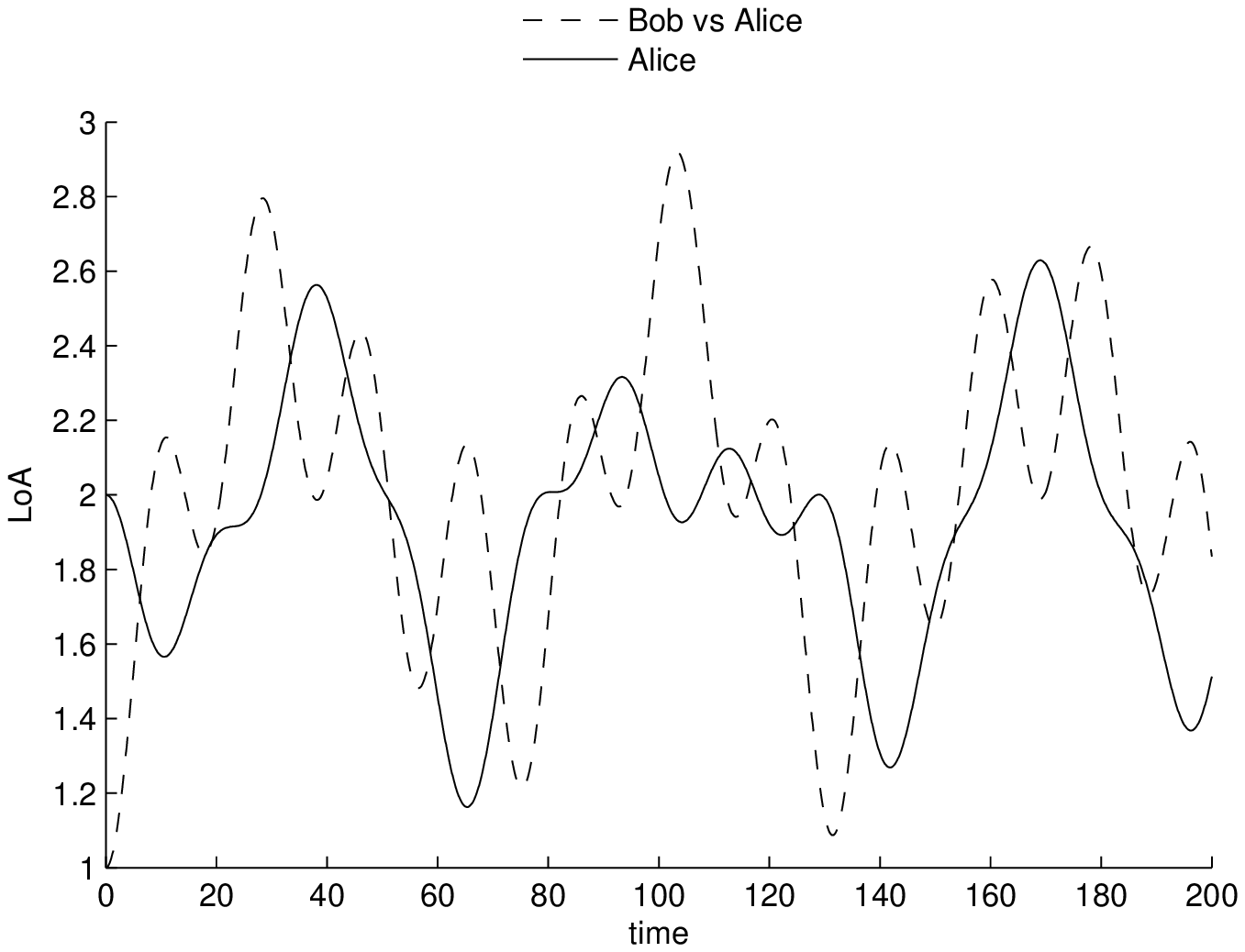}\hspace{8mm}
\includegraphics[width=0.47\textwidth] {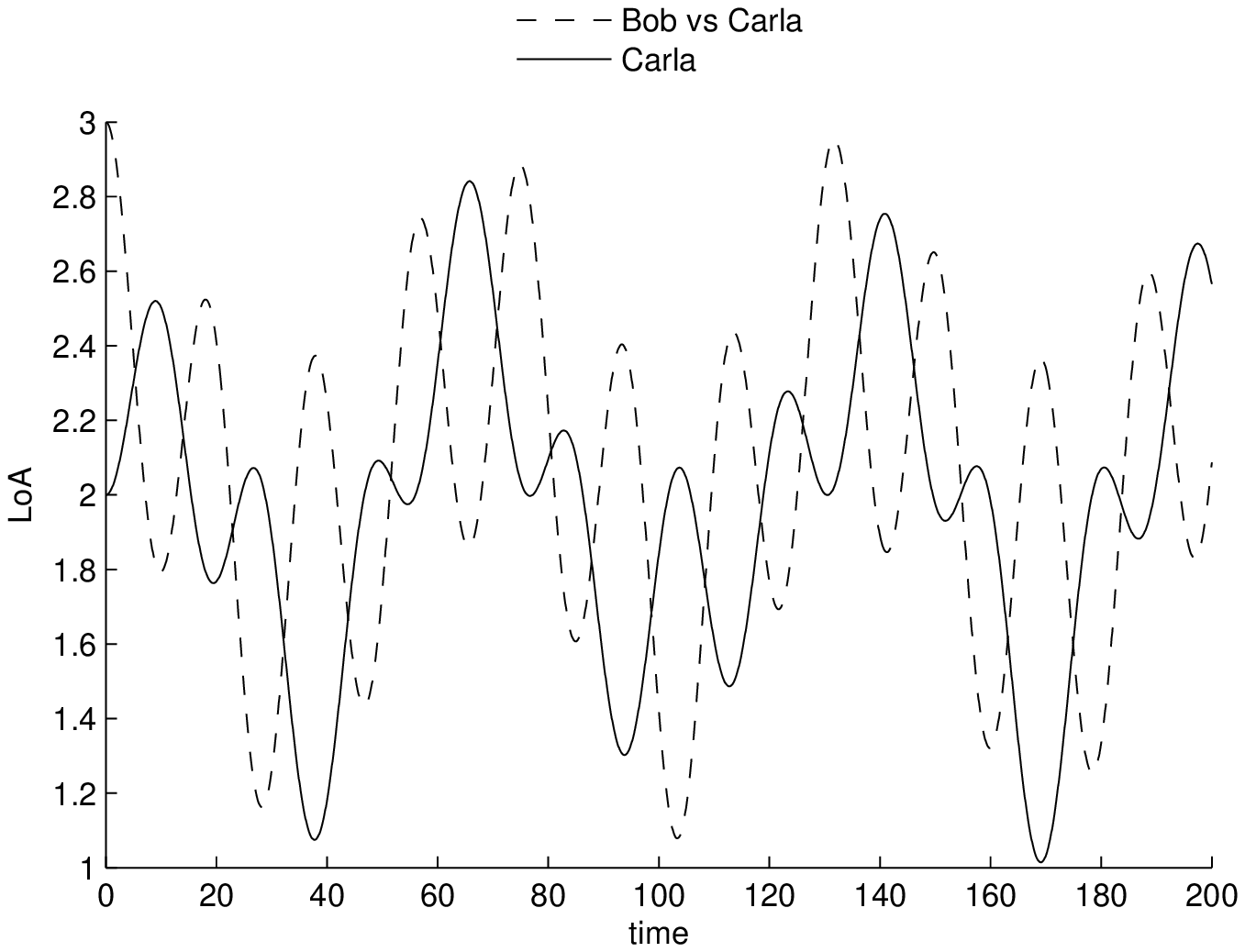}\hfill\\
\caption{\label{fig7}\footnotesize ($L_1=L_2=3$: Alice's and Bob's
LoA's vs. time with initial condition $(0,2)$ (left) and Carla's
and Bob's LoA's vs. time with initial condition $(2,3)$ (right). Quasiperiodic behaviours are observed.}
\end{center}
\end{figure}

The periodic behavior is clearly evident in both these plots. On
the contrary, in Figure~\ref{fig7}, we plot the solutions in the
quasiperiodic case for the same $L_1$, $L_2$, $n_{12}(0)$,
$n_{13}(0)$, $n_2(0)$ and $n_3(0)$ as above. The values of the
parameters are $\lambda_{12}=\frac{1}{10}$,
$\lambda_{13}=\frac{1}{8}$ and $\lambda_{1}\simeq 0.0889$, which
satisfy condition (\ref{33bis}) for $p=\sqrt{5}$ and $q=1$.

\subsection{Another generalization}
\label{sec:lovetriangle_sub1} A natural way to extend our previous
Hamiltonian consists, in view of what we have done in
Subsection~\ref{sec:firstmodel_sub1}, in introducing two different
parameters $M_\alpha$, $\alpha=12,13$, which are able to describe
the different (relative) reactions in the two interactions
Alice--Bob and Carla--Bob. We also assume that Bob is not very
interested in choosing  Carla rather than Alice, as long as one of
the two is attracted by him. For these reasons, the Hamiltonian
looks now like
\[
H=\lambda_{12}\left((a_{12}^\dagger)^{M_{12}}\,a_2+a_{12}^{M_{12}}\,a_2^\dagger\right)+\lambda_{13}\left((a_{13}^\dagger)^{M_{13}}\,a_3+a_{13}^{M_{12}}
\,a_3^\dagger\right)+
\lambda_{1}\left(a_{12}^\dagger\,a_{13}+a_{12}\,a_{13}^\dagger\right).
\]
Also in this case an integral of motion does exist, and looks like
\[
\tilde J:=N_{12}+N_{13}+M_{12}N_2+M_{13}N_3,
\]
which reduces to $J$ if $M_{12}=M_{13}=1$. The equations of motion
also extend those given in (\ref{32}), \emph{i.e.}, \bea\left\{
\begin{array}{ll} i\,\dot a_{12}(t)=\lambda_{12}\,M_{12} \,(a_{12}^\dagger(t))^{M_{12}-1}a_2(t)+\lambda_1\,a_{13}(t),\\
i\,\dot a_{13}(t)=\lambda_{13}\,M_{13} \,(a_{13}^\dagger(t))^{M_{13}-1} \,a_3(t)+\lambda_1\,a_{12}(t),\\
i\,\dot a_2(t)=\lambda_{12}\,a_{12}(t),\\
i\,\dot a_3(t)=\lambda_{13}\,a_{13}(t).
\end{array}\right.
\label{35} \ena System (\ref{35}) is nonlinear and can not be
solved analytically unless if $M_{12}=M_{13}=1$, as shown before.
Therefore, we adopt the same approach we have considered in our
first model.

In particular, we take $L_1=L_2=2$,  consider the orthonormal
basis $\F=\left\{\varphi_{ijkl}\right\}$, where the indices
$i,j,k,l$ run over the values 0,1,2 in such a way the sequence of
four--digit numbers $ijkl$ (in base-3 numeral system) is sorted in ascending
order, whereupon   $dim(\Hil_{eff})=(L_1+1)^2(L_2+1)^2=81$. In
such a situation, the unknowns $a_{12}(t)$, $a_{13}(t)$, $a_2(t)$
and $a_3(t)$ in the system (\ref{35}) are square matrices of
dimension 81, and their expression at $t=0$ are given by:
\[
\begin{aligned}
&a_{12}(0)= \left(
\begin{array}{lll}
\mathbf{0}_{27} & \mathbf{0}_{27} & \mathbf{0}_{27} \\
\Id_{27} & \mathbf{0}_{27} & \mathbf{0}_{27} \\
\mathbf{0}_{27} & \sqrt{2}\Id_{27} & \mathbf{0}_{27}
\end{array}
\right),\qquad &&a_{13}(0)= \left(
\begin{array}{lll}
\underline{x} & \mathbf{0}_{27} & \mathbf{0}_{27} \\
\mathbf{0}_{27} & \underline{x} & \mathbf{0}_{27} \\
\mathbf{0}_{27} & \mathbf{0}_{27} & \underline{x}
\end{array}
\right),\\
&a_{2}(0)= \left(
\begin{array}{lll}
\underline{y} & \mathbf{0}_{27} & \mathbf{0}_{27} \\
\mathbf{0}_{27} & \underline{y} & \mathbf{0}_{27} \\
\mathbf{0}_{27} & \mathbf{0}_{27} & \underline{y}
\end{array}
\right),\qquad &&a_{3}(0)= \left(
\begin{array}{lll}
\underline{z} & \mathbf{0}_{27} & \mathbf{0}_{27} \\
\mathbf{0}_{27} & \underline{z} & \mathbf{0}_{27} \\
\mathbf{0}_{27} & \mathbf{0}_{27} & \underline{z}
\end{array}
\right),
\end{aligned}
\]
where
\[
\begin{aligned}
&\underline{x} = \left(
\begin{array}{lll}
\mathbf{0}_9 & \mathbf{0}_9 & \mathbf{0}_9 \\
\Id_9 & \mathbf{0}_9 & \mathbf{0}_9 \\
\mathbf{0}_9 & \sqrt{2}\Id_9 & \mathbf{0}_9
\end{array}
\right),\qquad &&\underline{y}= \left(
\begin{array}{lll}
\underline{y}_1 & \mathbf{0}_9 & \mathbf{0}_9 \\
\mathbf{0}_9 & \underline{y}_1 & \mathbf{0}_9 \\
\mathbf{0}_9 & \mathbf{0}_9 & \underline{y}_1
\end{array}
\right),\\
&\underline{y}_1 = \left(
\begin{array}{lll}
\mathbf{0}_3 & \mathbf{0}_3 & \mathbf{0}_3 \\
\Id_3 & \mathbf{0}_3 & \mathbf{0}_3 \\
\mathbf{0}_3 & \sqrt{2}\Id_3 & \mathbf{0}_3
\end{array}
\right),\qquad &&\underline{z}= \left(
\begin{array}{lll}
\underline{z}_1 & \mathbf{0}_9 & \mathbf{0}_9 \\
\mathbf{0}_9 & \underline{z}_1 & \mathbf{0}_9 \\
\mathbf{0}_9 & \mathbf{0} & \underline{z}_1
\end{array}
\right),\\
&\underline{z}_1 = \left(
\begin{array}{lll}
\underline{z}_2 & \mathbf{0}_3 & \mathbf{0}_3 \\
\mathbf{0}_3 & \underline{z}_2 & \mathbf{0}_3 \\
\mathbf{0}_3 & \mathbf{0}_3 & \underline{z}_2
\end{array}
\right),\qquad &&\underline{z}_2 = \left(
\begin{array}{lll}
0 & 0 & 0 \\
1 & 0 & 0 \\
0 & \sqrt{2} & 0
\end{array}
\right).
\end{aligned}
\]

Moreover, in order to consider a situation not very far from a
linear one, we choose $M_{12}=1$, $M_{13}=2$, $\lambda_1=0.3$,
$\lambda_{12}=0.2$, $\lambda_{13}=0.007$.

\begin{figure}
\begin{center}
\includegraphics[width=0.7\textwidth]{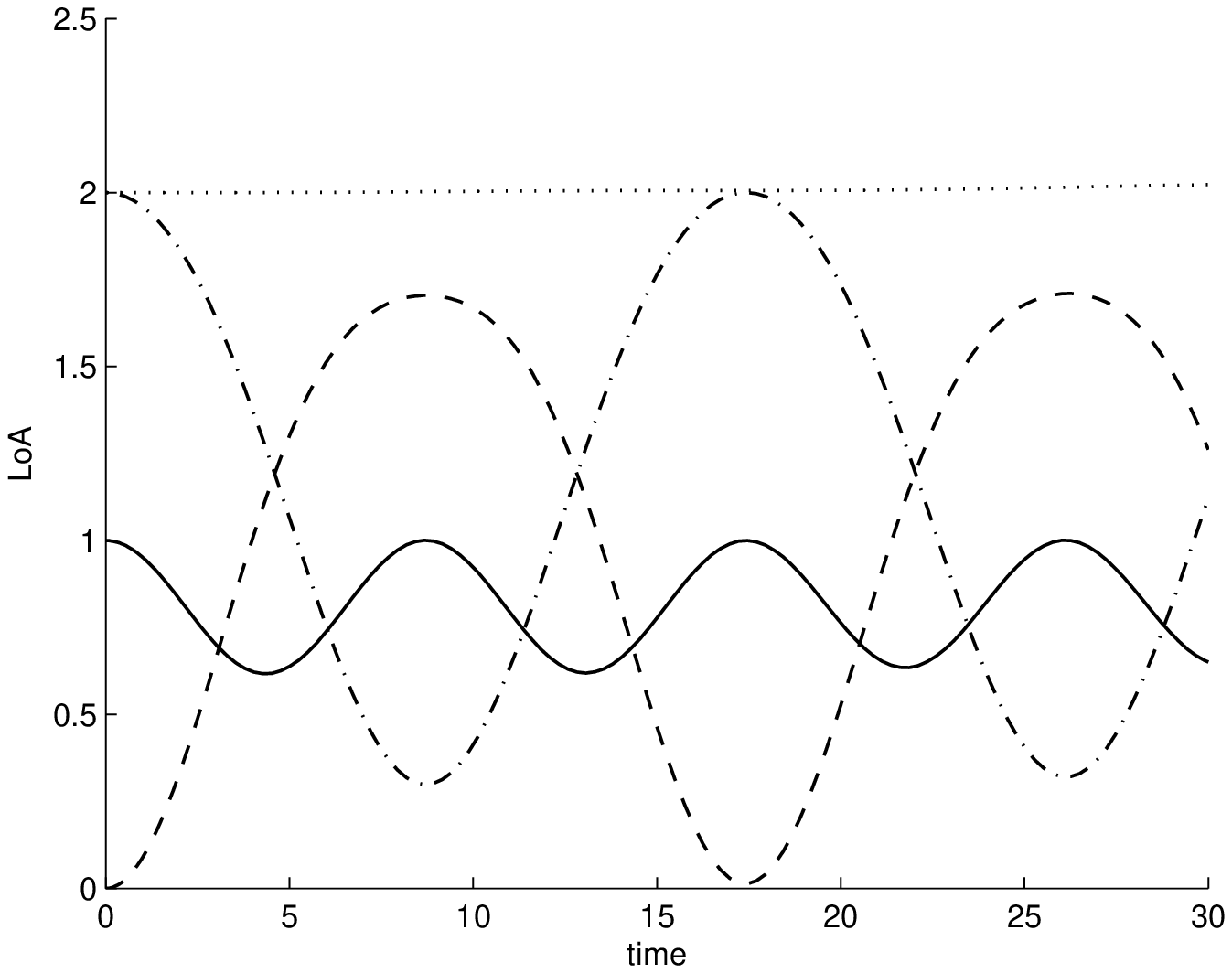}
\end{center}
\caption{\label{fig8}\footnotesize ($L_\alpha=2$, $M_{12}=1$, $M_{13}=2$
($\alpha=12,13,2,3$): LoA vs. time of: Bob vs. Alice (continuous
line), Bob vs. Carla (dashed line), Alice (dashed--dotted line),
Carla (dotted line) with initial condition $(1,0,2,2)$. Periodic behaviours are observed.}
\end{figure}

\begin{figure}
\begin{center}
\includegraphics[width=0.7\textwidth]{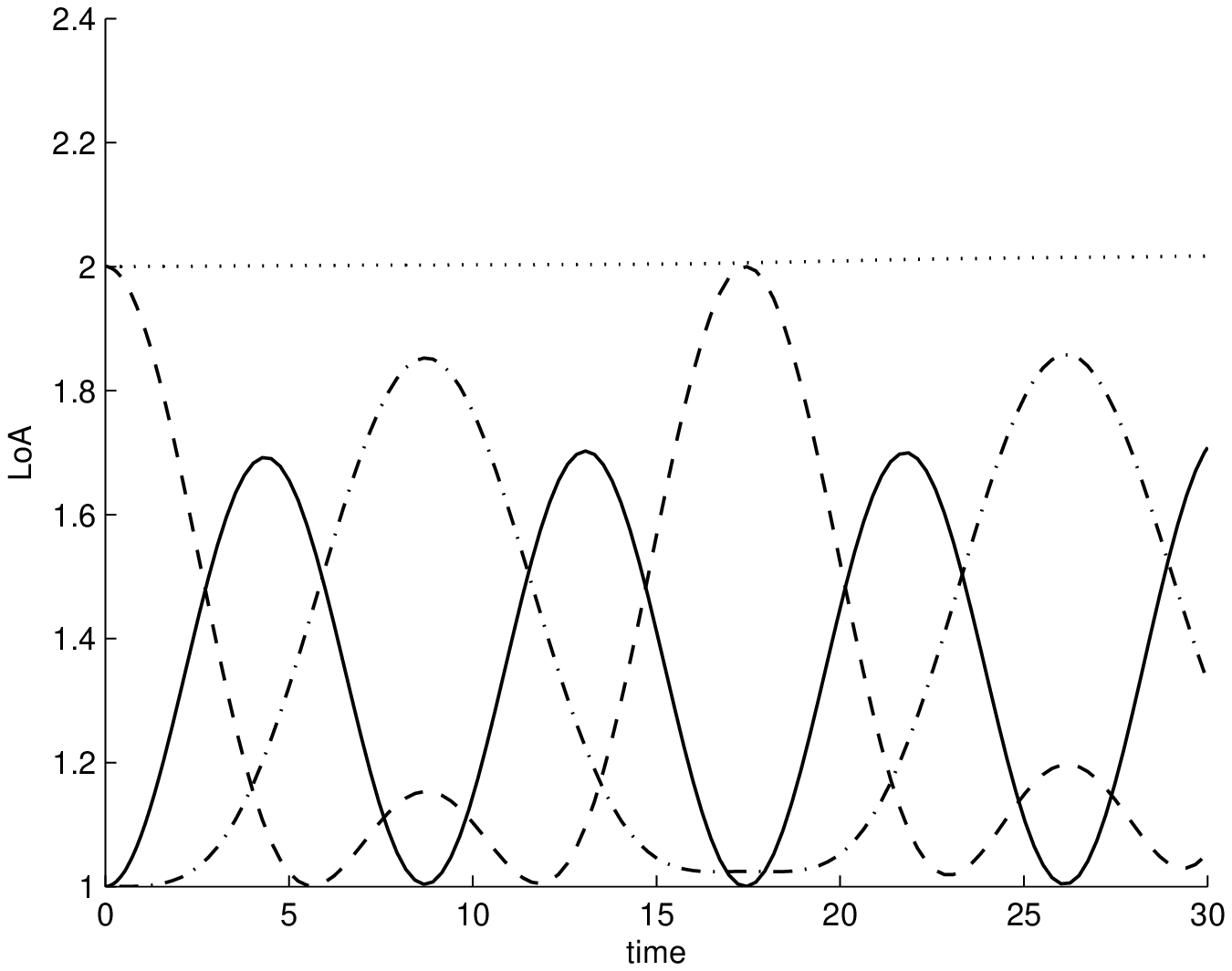}
\end{center}
\caption{\label{fig9}\footnotesize ($L_\alpha=2$, $M_{12}=1$, $M_{13}=2$
($\alpha=12,13,2,3$): LoA vs. time of: Bob vs. Alice (continuous
line), Bob vs. Carla (dashed line), Alice (dashed--dotted line),
Carla (dotted line) with initial condition $(1,2,1,2)$. Periodic behaviours are observed.}
\end{figure}

\begin{figure}
\begin{center}
\includegraphics[width=0.7\textwidth]{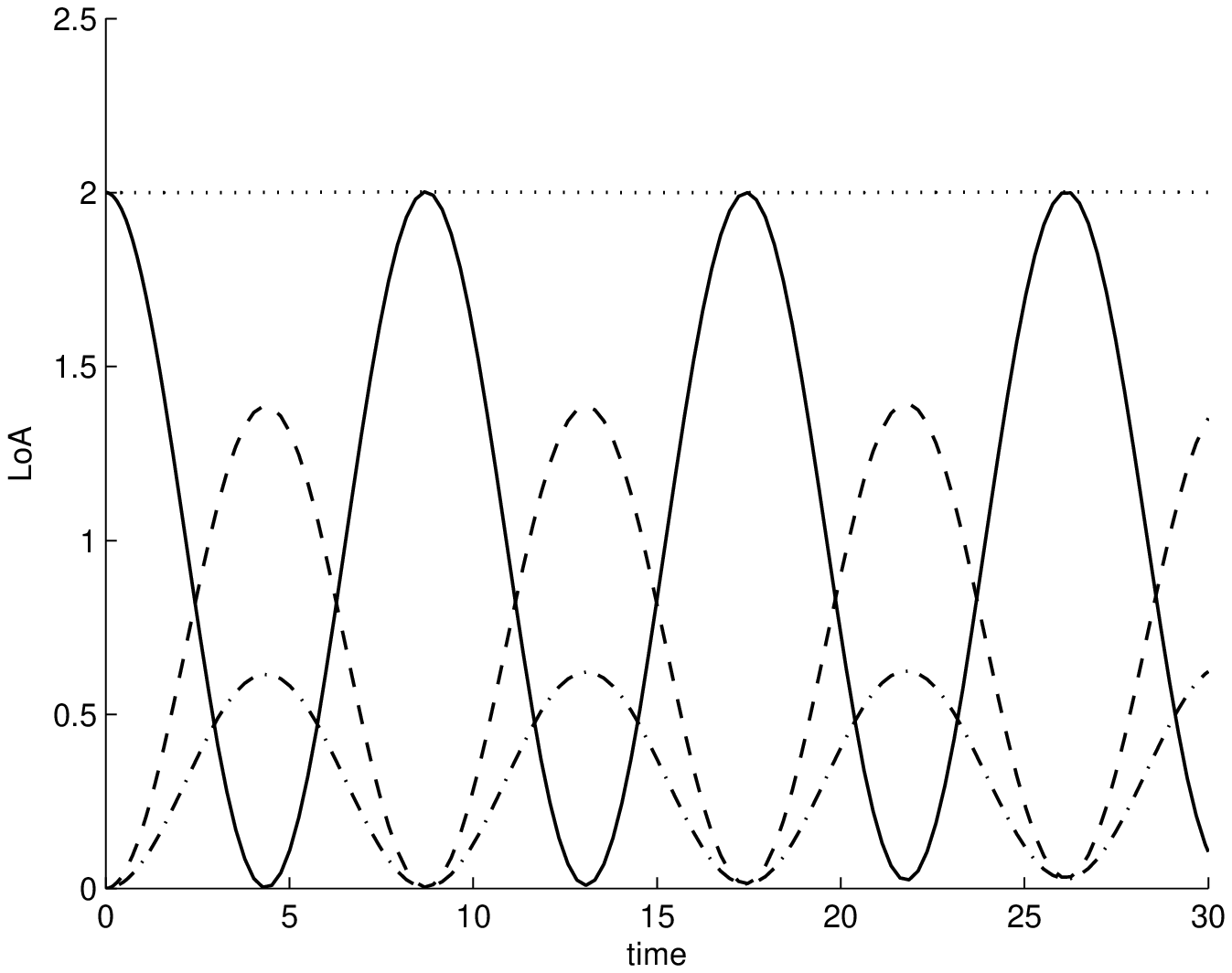}
\end{center}
\caption{\label{fig10}\footnotesize ($L_\alpha=2$, $M_{12}=1$, $M_{13}=2$
($\alpha=12,13,2,3$): LoA vs. time of: Bob vs. Alice (continuous
line), Bob vs. Carla (dashed line), Alice (dashed--dotted line),
Carla (dotted line) with initial condition $(2,0,0,2)$. Periodic behaviours are observed.}
\end{figure}

\begin{figure}
\begin{center}
\includegraphics[width=0.7\textwidth]{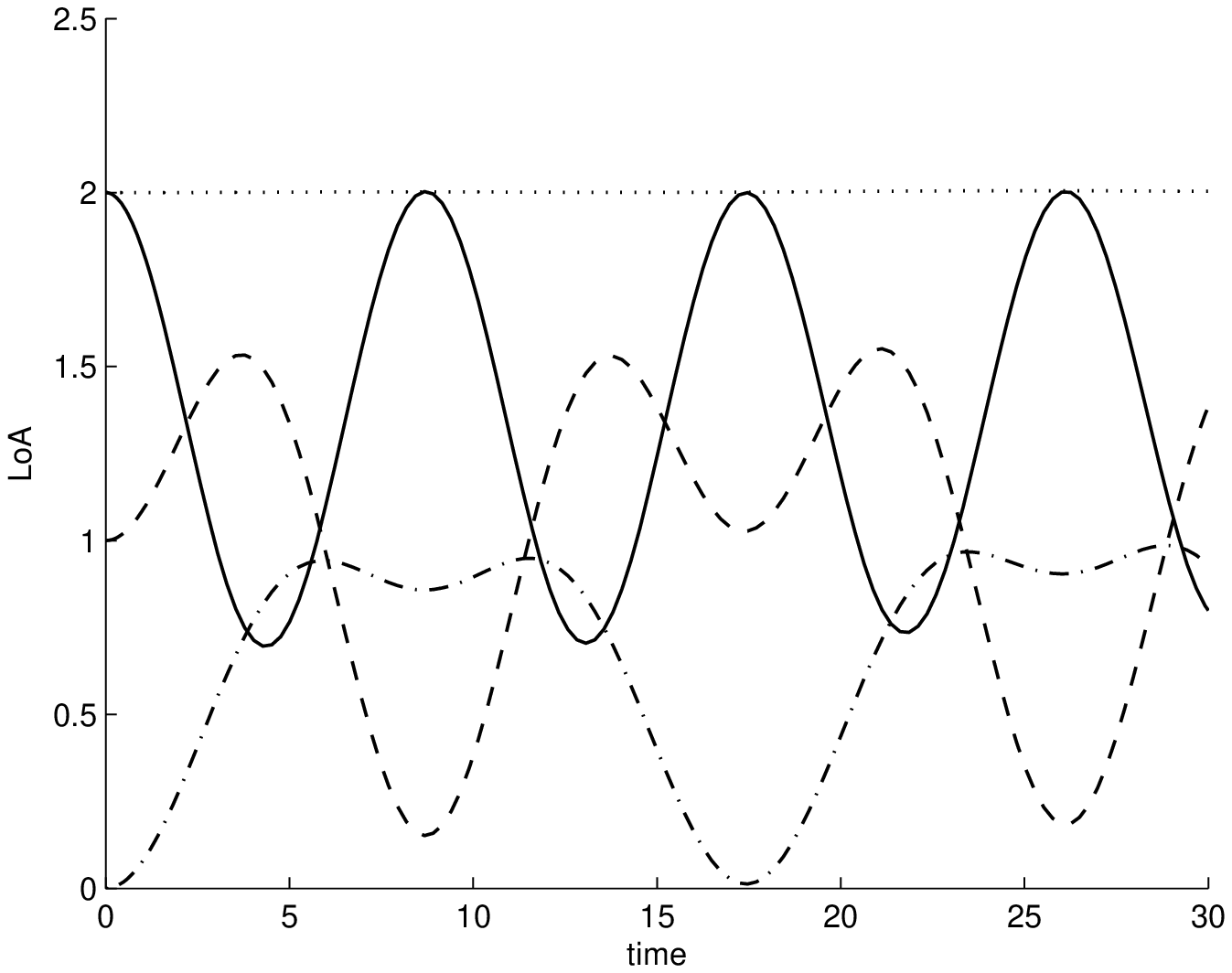}
\end{center}
\caption{\label{fig11}\footnotesize ($L_\alpha=2$, $M_{12}=1$, $M_{13}=2$
($\alpha=12,13,2,3$): LoA vs. time of: Bob vs. Alice (continuous
line), Bob vs. Carla (dashed line), Alice (dashed--dotted line),
Carla (dotted line) with initial condition $(2,1,0,2)$. Periodic behaviours are observed.}
\end{figure}

The numerical integration has been performed by using once again
the \hbox{ode113} routine of MATLAB$^\circledR$, and the existence
of the integral of motion has been used in order to check the
accuracy of the solutions.

The choice $\lambda_{13}=0.007$ has an immediate consequence,
which is clearly displayed by the plots in figures \ref{fig8}-\ref{fig11}. The time evolution of
$N_{3}(t)$ is very slow and essentially trivial in the time
interval where the system is numerically integrated: Carla keeps
her initial status. This does not imply, however, that Bob does
not change his LoA concerning Carla, as the figures \ref{fig8}-\ref{fig11} also clearly show.
Another interesting feature, which is again due to the nonlinear
aspect of the dynamics (and it was not present in the linear
model), is related to the fact that, for certain time intervals,
Bob's LoA for both Alice and Carla can increase. This is, in a
certain sense, unexpected: the Hamiltonian does not present any
explicit ingredient responsible for this feature, which, in our
opinion, appears just because of the nonlinearity of the
interaction. Finally, we also notice that the figures \ref{fig8}-\ref{fig11} display a
periodic behavior, even if no analytic proof of the existence of
this periodicity was produced, up to now.

It is worth  noticing that, contrarily to what happened in Section 2
(for certain initial conditions the dimension of $\Hil_{eff}$ needs
to be increased in order to capture the right dynamics), all the numerical results obtained here
suggest that, once the dimension of $\Hil_{eff}$ is fixed (and this is done when the initial conditions are given), the time evolution
of the various LoA's can be described inside that  $\Hil_{eff}$, \emph{i.e.},
there is no need to enlarge the dimension of the effective Hilbert space. This could be related to our choice of the parameters, which, as already stated, produces a system with \emph{small} nonlinear interactions. For this reason, we do not expect this to be a general feature of the model: on the contrary, we expect that, changing significantly the values of the parameters, the  need for increasing the dimensionality of $\Hil_{eff}$ will arise also here. Further analysis in this direction is in progress. 

\section{Conclusions}
\label{sec:conclusions} In this paper we have shown how to use
quantum mechanical tools in the analysis of dynamical systems
which model a two-- and a three--actor love relationship.
Depending on the parameters which describe the system, linear or
nonlinear differential equations are recovered, and exact or
numerical solutions are found. These solutions show that a
nontrivial dynamical behavior can be obtained, and a necessary and sufficient
condition for the periodicity or quasiperiodicity of the solution in the linear love triangle
model has been found, (\ref{33bis}).

It is worth  stressing that, despite the quantum framework adopted
in this paper, all the observables relevant for the description of
the models considered here commute among them and, for this
reason, can be measured simultaneously and no Heisenberg
uncertainty principle should be invoked for these operators.

A final comment is related to the oscillatory behavior recovered in
this paper for the different LoA's. Indeed, this is expected,
due to the hamiltonian approach considered here. In a paper
which is now in preparation, \cite{bagdecay}, one of us is
considering the possibility of describing, using the same
techniques, other processes in which some time decay may occur because of possible interactions of the lovers with the environment: this, we believe, could be a reasonable method to construct a more realistic model, where also different psychological mechanisms can be taken into account.

%\appendix
\renewcommand{\theequation}{A.\arabic{equation}}

%\section{\hspace{-.7cm}ppendix:  Few results on the number representation}
\section*{Appendix:  A few results on the occupation number representation}
We discuss here a few important facts in quantum mechanics and
second quantization, paying not much attention to mathematical
problems arising from the fact that the operators involved are
quite often unbounded. More details can be found, for instance, in
\cite{mer,reed}.

Let $\Hil$ be an Hilbert space, and $B(\Hil)$ the set of all the
bounded operators on $\Hil$.    Let $\ST$ be our physical system,
and $\A$ the set of all the operators useful for a complete
description of $\ST$, which includes the \emph{observables } of
$\ST$. For simplicity, it is convenient to assume that  $\A$
coincides with $B(\Hil)$ itself, even if this is not always
possible. This aspect,  related to the importance of some
unbounded operators within our scheme, will not be considered
here, even because the Hilbert space $\Hil_{eff}$ considered in
this paper is finite--dimensional, which implies that the
operators in $\A$ are bounded matrices. The description of the
time evolution of $\ST$ is related to a self--adjoint operator
$H=H^\dagger$ which is called the \emph{Hamiltonian} of $\ST$, and
which in standard quantum mechanics represents  the energy of
$\ST$. We will adopt here the so--called \emph{Heisenberg}
representation, in which the time evolution of an observable
$X\in\A$ is given by \be X(t)=e^{iHt}Xe^{-iHt}, \label{a1} \en or,
equivalently, by the solution of the differential equation \be
\frac{dX(t)}{dt}=ie^{iHt}[H,X]e^{-iHt}=i[H,X(t)],\label{a2} \en
where $[A,B]:=AB-BA$ is the \emph{commutator} between $A$ and $B$.
The time evolution defined in this way is usually a one--parameter
group of automorphisms of $\A$.

An operator $Z\in\A$ is a \emph{constant of motion} if it commutes
with $H$. Indeed, in this case, equation (\ref{a2}) implies that
$\dot Z(t)=0$, so that $Z(t)=Z$ for all $t$.

In our paper a special role is played by the so--called
\emph{canonical commutation relations} (CCR): we say that a set of
operators $\{a_\ell,\,a_\ell^\dagger, \ell=1,2,\ldots,L\}$ satisfy
the CCR if the conditions \be [a_\ell,a_n^\dagger]=\delta_{\ell
n}\Id,\hspace{8mm} [a_\ell,a_n]=[a_\ell^\dagger,a_n^\dagger]=0
\label{a3} \en hold true for all $\ell,n=1,2,\ldots,L$. Here,
$\Id$ is the identity operator. These operators, which are widely
analyzed in any textbook in quantum mechanics (see,  for instance,
\cite{mer}) are those which are used to describe $L$ different
\emph{modes} of bosons. From these operators we can construct
$\hat n_\ell=a_\ell^\dagger a_\ell$ and $\hat N=\sum_{\ell=1}^L
\hat n_\ell$, which are both self--adjoint. In particular, $\hat
n_\ell$ is the \emph{number operator} for the $\ell$-th mode,
while $\hat N$ is the \emph{number operator of $\ST$}.

The Hilbert space of our system is constructed as follows: we
introduce the \emph{vacuum} of the theory, that is a vector
$\varphi_0$ which is annihilated by all the operators $a_\ell$:
$a_\ell\varphi_0=0$ for all $\ell=1,2,\ldots,L$. Then we act on
$\varphi_0$ with the  operators $a_\ell^\dagger$ and their powers:
\be \varphi_{n_1,n_2,\ldots,n_L}:=\frac{1}{\sqrt{n_1!\,n_2!\ldots
n_L!}}(a_1^\dagger)^{n_1}(a_2^\dagger)^{n_2}\cdots
(a_L^\dagger)^{n_L}\varphi_0, \label{a4} \en $n_l=0,1,2,\ldots$
for all $l$. These vectors form an orthonormal set and are
eigenstates of both $\hat n_\ell$ and $\hat N$: $\hat
n_\ell\varphi_{n_1,n_2,\ldots,n_L}=n_\ell\varphi_{n_1,n_2,\ldots,n_L}$
and $\hat
N\varphi_{n_1,n_2,\ldots,n_L}=N\varphi_{n_1,n_2,\ldots,n_L}$,
where $N=\sum_{\ell=1}^Ln_\ell$. Moreover, using the  CCR we
deduce that $\hat
n_\ell\left(a_\ell\varphi_{n_1,n_2,\ldots,n_L}\right)=(n_\ell-1)(a_\ell\varphi_{n_1,n_2,\ldots,n_L})$
and $\hat
n_\ell\left(a_\ell^\dagger\varphi_{n_1,n_2,\ldots,n_L}\right)=(n_\ell+1)(a_l^\dagger\varphi_{n_1,n_2,\ldots,n_L})$,
for all $\ell$. For these reasons, the following interpretation is
given: if the $L$ different modes of bosons of $\ST$ are described
by the vector $\varphi_{n_1,n_2,\ldots,n_L}$, this implies that
$n_1$ bosons are in the first mode, $n_2$ in the second mode, and
so on. The operator $\hat n_\ell$ acts on
$\varphi_{n_1,n_2,\ldots,n_L}$ and returns $n_\ell$, which is
exactly the number of bosons in the $\ell$--th mode. The operator
$\hat N$ counts the total number of bosons. Moreover, the operator
$a_\ell$ destroys a boson in the $\ell$--th mode, while
$a_\ell^\dagger$ creates a boson in the same mode. This is why
$a_\ell$ and $a_\ell^\dagger$ are usually called the
\emph{annihilation} and the \emph{creation} operators.

The Hilbert space $\Hil$ is obtained by taking the closure of the
linear span of all these vectors.

The vector $\varphi_{n_1,n_2,\ldots,n_L}$ in (\ref{a4}) defines a
\emph{vector (or number) state } over the algebra $\A$  as \be
\omega_{n_1,n_2,\ldots,n_L}(X)=
\langle\varphi_{n_1,n_2,\ldots,n_L},X\varphi_{n_1,n_2,\ldots,n_L}\rangle,
\label{a5} \en where $\langle\,,\,\rangle$ is the scalar product
in the Hilbert space $\Hil$. As we have discussed in
\cite{bag1,bag2,bag3,bag4}, these states are used to
\emph{project} from quantum to classical dynamics and to fix the
initial conditions of the system considered.

\section*{Acknowledgments}

 The authors wish to express their gratitude to the referees for their suggestions, which improved significantly the paper. This work has been financially supported in part by G.N.F.M. of
I.N.d.A.M., and by local Research Projects of the Universities of
Messina and Palermo.

\end{document}